\documentclass[aps,pra,twocolumn,superscriptaddress,10pt,showpacs]{revtex4-1}

\usepackage{multirow}
\usepackage[utf8x]{inputenc}
\usepackage{graphicx}
\usepackage{amsmath,amssymb}
\usepackage{datetime}
\usepackage{color}
\usepackage[svgnames]{xcolor}
\usepackage{array}
\usepackage[
pdftitle={Quantum Simulation Meets Nonequilibrium Dynamical Mean Field Theory: Exploring the Periodically Driven, Strongly Correlated Fermi-Hubbard Model},    % title
pdfauthor={Kilian Sandholzer, Yuta Murakami, Frederik Goerg, Joaquin Minguzzi, Michael Messer, Remi Desbuquois, Martin Eckstein, Philipp Werner, Tilman Esslinger},     % author
pdfsubject={Optical Lattice},   % subject of the document
pdfcreator={Kilian Sandholzer, Yuta Murakami, Frederik Goerg, Joaquin Minguzzi, Michael Messer, Remi Desbuquois, Martin Eckstein, Philipp Werner, Tilman Esslinger},   % creator of the document
pdfkeywords={Optical Lattice } {Driven Fermi-Hubbard Model } {Ultracold Fermions } {Floquet }, % list of keywords
colorlinks=True,linkcolor=DarkSlateBlue,citecolor=DarkBlue,urlcolor=DarkBlue,
	pdfstartview=FitH,bookmarks=False,pdfpagemode=UseNone
]{hyperref}

\begin{document}

\title{Quantum Simulation Meets Nonequilibrium Dynamical Mean Field Theory: Exploring the Periodically Driven, Strongly Correlated Fermi-Hubbard Model}

\author{Kilian Sandholzer}
\affiliation{Institute for Quantum Electronics, ETH Zurich, 8093 Zurich, Switzerland}
\author{Yuta Murakami}
\affiliation{Department of Physics, University of Fribourg, 1700 Fribourg, Switzerland}
\author{Frederik G\"org}
\author{Joaqu\'in Minguzzi}
\author{Michael Messer}
\author{R\'emi Desbuquois}
\affiliation{Institute for Quantum Electronics, ETH Zurich, 8093 Zurich, Switzerland}
\author{Martin Eckstein}
\affiliation{Department of Physics, University of Erlangen-N\"urnberg, 91058 Erlangen, Germany}
\author{Philipp Werner}
\affiliation{Department of Physics, University of Fribourg, 1700 Fribourg, Switzerland}
\author{Tilman Esslinger}
\affiliation{Institute for Quantum Electronics, ETH Zurich, 8093 Zurich, Switzerland}
\date{\today}

\begin{abstract}
We perform an ab-initio comparison between nonequilibrium dynamical mean-field theory and optical lattice experiments by studying the time evolution of double occupations in the periodically driven Fermi-Hubbard model. 
For off-resonant driving, the range of validity of a description in terms of an effective static Hamiltonian is determined and its breakdown due to energy absorption close to resonance is demonstrated. For near-resonant driving, we investigate the response to a change in driving amplitude and discover an asymmetric excitation spectrum with respect to the detuning. 
In general, we find good agreement between experiment and theory, which cross-validates the experimental and numerical approaches in a strongly-correlated nonequilibrium system.
\end{abstract}

\maketitle

Quantum simulation exploits the high degree of control over a quantum system, such as ultracold atoms, to explore the complexity of many-body physics \cite{Feynman1982,Georgescu2014,Bloch2012,Tarruell2018}. 
To gain reliable insights from this approach it is important to benchmark the simulator against numerical or analytical methods. Extensive comparisons have been performed for static systems, such as the Fermi-Hubbard model, which captures essential effects of electronic correlations in solids \cite{Jordens2010,Sciolla2013,Imriska2014,Golubeva2015,Hart2015,Cocchi2016,Mazurenko2017}. 
A new frontier in many-body physics is the study of driven systems, such as high-temperature superconductors exposed to intense laser fields \cite{Nicoletti2016,Mitrano2016} or cold atoms in topologically non-trivial band structures \cite{Goldman2016,Eckardt2017}.
It is indeed a considerable challenge to understand the consequences of periodic driving, often referred to as Floquet engineering, in correlated lattice models \cite{Oka2018,Eckardt2017}.
An important question is to what extent the properties of the driven system can be captured by an effective static description (Floquet Hamiltonian) despite its nonequilibrium nature.
We address this subject by studying the driven Fermi-Hubbard model in the experimental setting of an optical lattice and directly compare the results to nonequilibrium dynamical mean field theory (DMFT). 
 
Effective Floquet Hamiltonians can be derived from high-frequency expansions or time-dependent Schrieffer-Wolff transformations \cite{Goldman2014,Bukov2015b,Mikami2016,Bukov2016b}.
It is expected that these effective models describe the dynamics and thermodynamics of the many-body system under certain conditions. 
Nevertheless, the real Floquet-engineered state may be characterized by non-thermal energy distributions induced by switch-on procedures or energy absorption from the periodic drive  \cite{Bukov2015b,Eckstein2017,Singh2018,Mori2018,Moessner2017,Herrmann2018} and higher order corrections.
We use ultracold fermions in a brick wall lattice to avoid unwanted excitation processes induced by the driving  \cite{Messer2018} that otherwise lead to heating of the interacting system \cite{Weinberg2015,Strater2016,Reitter2017,Lellouch2017,Nager2018,Boulier2018}.
A theoretical formalism which captures the full time evolution is nonequilibrium DMFT \cite{Georges1996,Freericks2006,Aoki2014}.
This method has been used to study a broad range of nonequilibrium setups in single-band \cite{Aoki2014} and multi-band \cite{Strand2017} Hubbard models, and to interpret pump-probe experiments on correlated solids at a qualitative level \cite{Ligges2018}. 
However, there have been only few attempts to test the accuracy of this method for the nonequilibrium dynamics in finite-dimensional lattices \cite{Tsuji2014} and there has so far been no ab-initio comparison to experiments.

\begin{figure}
    \includegraphics[width=\columnwidth]{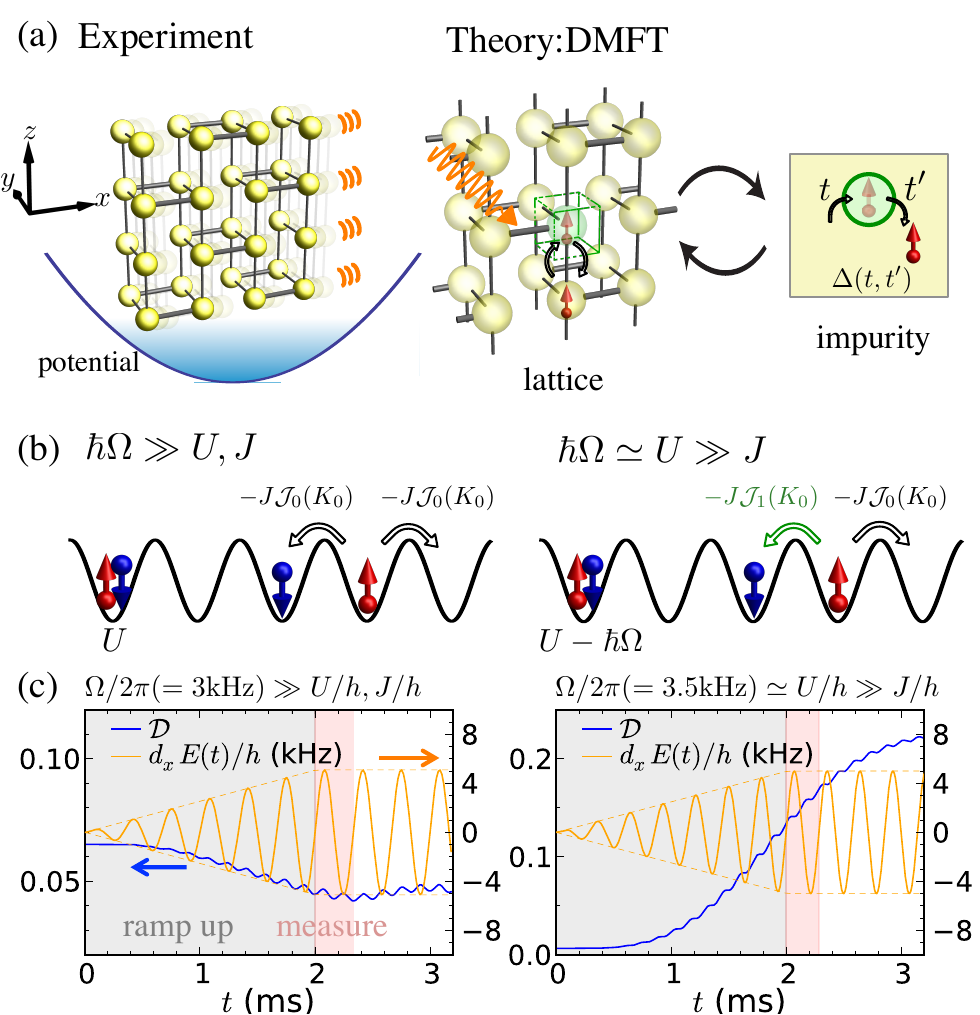}
    \caption{(a) Experiment: Three-dimensional brick wall structure in a trapping potential. 
    The driving is applied in the $x$ direction. 
    Theory: DMFT mapping of the lattice system to an effective impurity problem. 
    It is characterized by the hybridization function $\Delta(t,t')$, which mimics the hopping of particles to neighboring sites in the lattice system. (b) Schematic illustration of the different effective Hamiltonians. 
    In the off-resonant regime ($\hbar \Omega \gg U,J$), the interaction $U$ is unaffected while the hopping parameter $J$ is renormalized. In the near-resonant regime ($\hbar \Omega \approx U \gg J$), the interactions are reduced to $U-\hbar\Omega$ and the hopping parameter depends on whether or not the tunneling process changes the number of doubly occupied sites. 
    (c) DMFT simulations of the double occupation $\mathcal{D}$ in the off-resonant ($\Omega/2\pi = 3\ \mathrm{kHz}, U/h = 750\ \mathrm{Hz}, J_x = 200\ \mathrm{Hz}, J_y=J_z = 40\ \mathrm{Hz}$) and near-resonant ($\Omega/2\pi = U/h = 3.5\ \mathrm{kHz}, J_x = 200\ \mathrm{Hz}, J_y=J_z = 100\ \mathrm{Hz}$) regimes. 
    As in the experiment, the driving field $E(t)$ is ramped up linearly during a period $t_{\rm ramp}$ and $\mathcal{D}$ is measured just after the ramp and averaged over one period of the excitation ($T=\frac{2\pi}{\Omega}$).}\label{fig1}
\end{figure}

We investigate the driven Fermi-Hubbard model 
\begin{equation}
\hat{H}(t)=-\sum_{\substack{\left\langle \mathbf{i},\mathbf{j}\right\rangle, \sigma}} J_{\mathbf{ij}} c^{\dagger}_{\mathbf{i}\sigma}c_{\mathbf{j}\sigma}+U \sum_\mathbf{i} \hat{n}_{\mathbf{i}\uparrow}\hat{n}_{\mathbf{i}\downarrow} + E(t) \sum_{\mathbf{i},\sigma} x_{\mathbf{i}} \hat{n}_{\mathbf{i}\sigma}
\label{drivenFHM}
\end{equation}
on a three-dimensional brick wall lattice structure (Fig.~\ref{fig1}a). 
Here, $\hat{c}^{\dagger}_{\mathbf{i}\sigma}$ and $\hat{n}_{\mathbf{i}\sigma}$ are the fermionic creation and number operators at site $\mathbf{i}=(i_x,i_y,i_z)$ in spin-state $\sigma=\uparrow,\downarrow$, respectively. 
The nearest neighbor $\left\langle \mathbf{i},\mathbf{j}\right\rangle$ tunneling rate is denoted by $J_{\mathbf{ij}}$, the onsite interaction by $U$ and the time-periodic driving field in the $x$ direction by $E(t)$, with $x_{\mathbf{i}}=\left\langle \hat{x}\right\rangle_{\mathbf{i}}$ the $x$-position of the Wannier function on site $\mathbf{i}$. 
Our study covers the off-resonant and near-resonant driving regimes, which are described by different effective Hamiltonians, see Fig.~\ref{fig1}b. 

Experimentally, the model is implemented using a degenerate fermionic $^{40}\text{K}$ cloud with $N=35(3) \times 10^3$ atoms \cite{si4} loaded into a three-dimensional optical lattice with a brick wall geometry \cite{Tarruell2012}.
Two equally populated magnetic sublevels of the $F=9/2$ hyperfine manifold mimic the interacting spin up and down particles moving in the lowest band of the lattice.
The time-periodic field in Eq.~\eqref{drivenFHM} is realized with a piezoelectric actuator moving the retroreflecting mirror of the lattice such that the $x$ position of the lattice is sinusoidally modulated \cite{si4}. 
It can be written as $E(t)= m A \Omega^2 \sin(\Omega t)$, where $A$ is the displacement amplitude, $m$ the mass of the atoms and $\Omega$ the angular driving frequency.
A crossed dipole trap forms an overall harmonic confinement on top of the periodic potential generated by the lattice beams. The resulting atomic density distribution can be estimated in the loading lattice for an independently determined entropy \cite{si4}.

On the theory side the same model is studied using nonequilibrium DMFT \cite{Georges1996,Freericks2006,Aoki2014} (for details of the implementation, see \cite{si1}). 
DMFT is based on a self-consistent mapping to a quantum impurity model (Fig.~\ref{fig1}a) and a local self-energy approximation, which becomes exact in the limit of infinite dimensions \cite{Metzner1989,Mueller-Hartmann1989}. 
The periodic driving is incorporated by a Peierls factor in the hopping terms \cite{Aoki2014}. 
To solve the impurity problem, we use the non-crossing (NCA) and one-crossing (OCA) approximations \cite{Keiter1971,Pruschke1989, Eckstein2010}. It turns out that NCA is sufficient to describe the system in the present study \cite{si1}.
The local density approximation (LDA) is employed to take into account the inhomogeneity of the cold atoms system, i.e., we simulate the dynamics of homogeneous systems with different fillings and compute the average over the experimentally determined density distribution \cite{si1,si4}. 
The comparison between theory and experiment is thus limited to timescales which are short enough that there is no significant redistribution of atoms within the trap \cite{Messer2018}. 

The many-body dynamics can be captured by measuring the fraction of atoms on doubly occupied sites $\mathcal{D}=2/N \sum_{\mathbf{i}} \left\langle \hat{n}_{\mathbf{i}\uparrow}\hat{n}_{\mathbf{i}\downarrow}\right\rangle$ \cite{si1,si4}.
This observable indicates how the nature of the state changes when the effective on-site interaction changes or the system is driven in or out of strongly correlated regions. The value of ${\mathcal D}$ is averaged over the spatially inhomogeneous system and one driving cycle to distinguish the effective dynamics from micromotion \cite{Desbuquois2017,Goerg2018}. 
Theoretical plots illustrating the full time evolution of ${\mathcal D}$ and the measurement protocol are shown in Fig.~\ref{fig1}c. In addition, DMFT calculations allow us to extract the local single-particle spectral function $A(\omega)$ and its particle (hole) occupation $N(\omega)$ ($\bar{N}(\omega)$) to investigate the driving induced couplings between many-body states \cite{si1}.

In the off-resonant case $\hbar\Omega \gg U, W$, with $W=2J_x+4(J_y+J_z)$ denoting the free-fermion bandwidth, a high-frequency expansion to lowest order yields the effective Hamiltonian \cite{Dunlap1986,Eckardt2005a}
\begin{align}
\hat{H}^{\mathrm{eff}}_{\mathrm{off-res}}=
& - J_{x} \mathcal{J}_0(K_0) \sum_{\langle \mathrm{\textbf{i,j}} \rangle_x , \sigma} \hat{c}^{\dagger}_{\textbf{i} \sigma}\hat{c}_{\textbf{j} \sigma}\nonumber \\ 
&-J_{y,z} \sum_{\langle \textbf{i,j} \rangle_{y,z} , \sigma} \hat{c}^{\dagger}_{\textbf{i} \sigma}\hat{c}_{\textbf{j} \sigma} 
 + U \sum_{\textbf{i}} \hat{n}_{\textbf{i} \downarrow}\hat{n}_{\textbf{i} \uparrow}.
\label{Heff_off_res}
\end{align}
This corresponds to a static Hubbard model with hopping in the $x$ direction renormalized by the zeroth-order Bessel function $\mathcal{J}_0(K_0)$ \cite{Lignier2007,Zenesini2009} which depends on the dimensionless driving amplitude $K_0=mA\Omega d_x/\hbar$; $d_x$ denotes the distance of two neighboring sites in the $x$ direction. 
If we lower the driving frequency, higher order corrections to Eq.~\eqref{Heff_off_res} have to be taken into account and reliable information on the evolution of the state can only be obtained by the combination of quantum simulations and time-dependent DMFT calculations.

\begin{figure}[t]
    \includegraphics[width=\columnwidth]{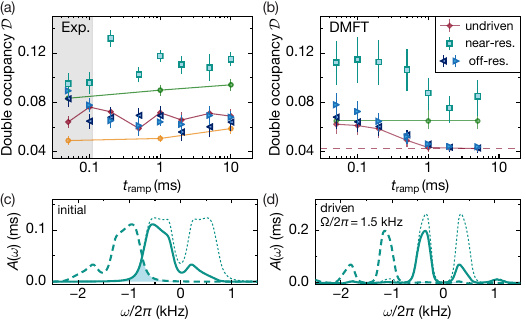}    
    \caption{Off-resonant driving in the Fermi-Hubbard model at $U/h=0.75(3)$~kHz and $W/h=0.71(7)$~kHz. 
    Experimental (a) and DMFT (b) data display $\mathcal{D}$ measured after ramping up the drive on different timescales. 
    The legend applies to both plots. 
    In the off-resonant case (triangle left: $\Omega/2\pi=5$~kHz, triangle right: $\Omega/2\pi=3$~kHz) $\mathcal{D}$ is suppressed, whereas an increase is observed when $\Omega/2\pi=1.5\ \mathrm{kHz} \simeq (U+W)/h$ (squares). 
    Solid lines show data taken in an undriven system. The upper and lower lines are reference values for holding in the initial ($J_x/h=193(34)$~Hz) and final lattice ($J_x^{\text{eff}}/h=81(13)$~Hz). 
    The red line displays data taken after ramping the lattice depth from $J_x$ to $J_x^{\text{eff}}$ to mimic the driven data. 
    The dashed red line indicates the saturation value reached at $t_\text{ramp}=5$~ms. 
    Data in the shaded area of (a) is influenced by residual dynamics during detection and the finite bandwidth of the piezoelectric actuator. 
    Error bars in (a) denote the standard error for 5 measurements and in (b) reflect the uncertainty of the entropy estimation in the experiment \cite{si4}. 
    Panel (c) shows the local single-particle spectral function $A(\omega)$ (thin dashed) and its occupation $N(\omega)$ (solid) at $T/J_x = 1.21$ in equilibrium at half-filling. 
	The shaded area indicates the overlap between $N(\omega)$ and the hole occupation $\bar{N}(\omega-\Omega)$ shifted by the driving frequency $\Omega/2\pi=1.5$~kHz (dashed), corresponding to possible direct excitations. 
	In (d) we plot the nonequilibrium spectra after ramping up the drive in 5~ms.}\label{fig2}
\end{figure}

For $U/W=1.1(1)$ we compare experimental (Fig.~\ref{fig2}a) and theoretical (Fig.~\ref{fig2}b) data for different drive frequencies in the off-resonant regime to first validate the effective Hamiltonian description according to Eq.~\eqref{Heff_off_res}. 
We prepare an initial state with $\mathcal{D} = 0.083(5)$ \cite{si2,si4} and ramp up the modulation in different times $t_{\text{ramp}}$ to the final strength ($K_0 =1.68(2)$) before $\mathcal{D}$ is measured.
If the renormalization to $J_{\text{eff}}$ is dominant, $\mathcal{D}$ is expected to be suppressed because $U/J_{\text{eff}}$ increases.
This is verified by data taken in static lattices at the same entropy, with the hopping set once to the preparation parameter ($J_x$, green) and once to the effective value ($J_x^\text{eff}=J_x\mathcal{J}_0(K_0)$, orange). 
Theoretically, the same reference line for the preparation lattice is calculated (green) and for $J_x/h=81(12)$~Hz the equilibration value of $\mathcal{D}$ is estimated by the adiabatic ramp of the hopping from $J_x/h=193(32)$~Hz \cite{si2}. 
The cloud is loaded into a shallow lattice to achieve adiabatic preparation. However, in order to resolve the dynamics in the experiment, the tunneling energies in the starting lattice of the experiment are reduced. 
This leads to a gradual increase in $\mathcal{D}$ associated with the induced global density redistribution \cite{si4}.

To test the validity of the effective Hamiltonian \eqref{Heff_off_res} we also simulate undriven systems in which the hopping amplitude is changed in time by a protocol which mimics the ramp-up of the driving amplitude $K_0$ in the effective Hamiltonian (red lines) \cite{Goerg2018,Messer2018}.
For the large (off-resonant) driving frequencies, both theoretical and experimental results follow the trend of the effective Hamiltonian dynamics. The theoretical data clearly identifies adiabatic time scales above 1~ms for reaching the equilibrium reference value which is consistent with the experimental data, although the latter are not fully conclusive due to the large scatter.
In addition, the theoretical results show that the effective description is valid even when $t_{\rm ramp}$ is comparable to a single cycle ($(2\pi)/\Omega = 0.2$ to $0.33$~ms) of the modulation \cite{Eckstein2017,si2}.

By moving the drive frequency closer to resonances with the onsite interaction $U$ (see also \cite{si4}), we explore for which frequencies Eq.~\eqref{Heff_off_res} still provides a good description of the system \cite{si2}.
At $\Omega/2\pi$ = 1.5 kHz the frequency is larger than $U$ and $W$ but comparable to $U+W$, which is the naively expected maximum energy of a double occupation excitation in the system. 
In this non-trivial regime both theory and experiment consistently predict a breakdown of the effective description.

Here $\mathcal{D}$ are created at short ramp times before decreasing again for longer times. 
Experimentally, times below 0.1~ms (shaded area) are difficult to access because of the finite bandwidth of the piezoelectric actuator and the detection time. 
From the theoretically obtained local single-particle spectral function one can see that direct excitations across the gap are possible because the bandwidth is broadened by the interaction (Fig.~\ref{fig2}c). 
This is most pronounced at short times as the effective bandwidth decreases due to the driving at longer times (Fig.~\ref{fig2}d). 
Interestingly, $\mathcal{D}$ does not decrease to the same values as for higher driving frequencies beyond $1$~ms. 
Despite a very similar final effective Hamiltonian, the final state is very different depending on the energy absorbed. 
This can be seen as non-adiabatic behaviour which was confirmed by further studies in the off- and near-resonant regime \cite{si2,si3,si4}. 
Since the number of coupled states changes rapidly with driving frequency, $\mathcal{D}$ is very sensitive to $\Omega$ in this regime. 
We attribute the remaining deviations in the values of $\mathcal{D}$ between the ab-initio calculations and the experimental values to the systematic uncertainties on the input temperature and density profiles provided by the experiment \cite{si2,si4}.

A particularly appealing feature of Floquet engineering is the possible creation of effective Hamiltonians with terms which are difficult to realize in static systems. An example in the strong coupling regime is the near-resonantly driven system ($J \ll U \simeq \hbar \Omega$), for which the effective Hamiltonian becomes \cite{Bukov2016b,Bermudez2015,Itin2015,Goerg2018}
\begin{align}
&\hat{H}^{\mathrm{eff}}_{\mathrm{res}} 
= - J_x \sum_{\textbf{i} \in \mathcal{A} , \sigma\atop \textbf{j}=\textbf{i}+\textbf{e}_x} \left[\left( \mathcal{J}_{0}(K_0) \hat{a}_{\textbf{ij}\overline{\sigma}} + \mathcal{J}_{1}(K_0) \hat{b}_{\textbf{ij}\overline{\sigma}} \right)
 \hat{c}^{\dagger}_{\textbf{i} \sigma}\hat{c}_{\textbf{j} \sigma}\right. \nonumber \\ 
&\left. +\mathrm{H.c.}\right] -J_{y,z} \sum_{\langle \textbf{i,j} \rangle_{y,z} , \sigma} \hat{a}_{\textbf{ij},\overline{\sigma}} \hat{c}^{\dagger}_{\textbf{i} \sigma}\hat{c}_{\textbf{j} \sigma} + \left(U - \hbar \Omega \right) \sum_{\textbf{i}} \hat{n}_{\textbf{i}\downarrow}\hat{n}_{\textbf{i} \uparrow},
\label{Heff_res}
\end{align}
as illustrated in the right panel of Fig.~\ref{fig1}b.
In comparison to the static Hubbard model, the interaction is modified to the detuning from the drive $U^{\mathrm{eff}}=U - \hbar \Omega$. 
The tunneling processes can be separated into two classes: (i) single particle tunneling processes which keep the number of double occupancies constant $\left[\hat{a}_{\textbf{ij} \overline{\sigma}}=(1-\hat{n}_{\textbf{i} \overline{\sigma}})(1-\hat{n}_{\textbf{j} \overline{\sigma}})+\hat{n}_{\textbf{i} \overline{\sigma}}\hat{n}_{\textbf{j} \overline{\sigma}} \, \text{and}\, \overline{\uparrow} = \downarrow \right]$, such that the interaction energy is irrelevant, and (ii) tunneling processes which increase or decrease the double occupancy by one unit $[\hat{b}_{\textbf{ij} \overline{\sigma}}= -(1-\hat{n}_{\textbf{i} \overline{\sigma}} ) \hat{n}_{\textbf{j} \overline{\sigma}}+\hat{n}_{\textbf{i} \overline{\sigma}}(1-\hat{n}_{\textbf{j} \overline{\sigma}})]$. 
Since one opposite spin particle is involved in the latter processes, these are density-dependent hoppings \cite{Ma2011, Meinert2016, Desbuquois2017, Xu2018, Goerg2018} which make $\hat{H}^{\mathrm{eff}}_{\mathrm{res}} $ fundamentally different from a static Hubbard model.

\begin{figure}
    \includegraphics[width=\columnwidth]{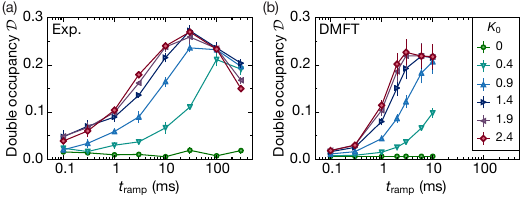}
    \caption{Resonant driving in the Fermi-Hubbard model for $J_{x,y,z}/h=[200(50),100(10),100(10)]$~Hz. 
    Experimentally measured (a) and theoretically simulated (b) $\mathcal{D}$ for different ramp times and driving strengths at resonance ($\Omega/(2\pi)=3.5$ kHz, $U/h=3.5(1)$ kHz).
	Dynamics beyond $t_{\mathrm{ramp}}=10$~ms are influenced by trap effects and for $t_{\mathrm{ramp}}=$300~ms by heating \cite{si4} and is not considered in DMFT. 
	Error bars are the same as in Fig.~\ref{fig2}.}
\label{fig3}
\end{figure}

In one set of measurements (Fig.~\ref{fig3}a (experiment) and \ref{fig3}b (theory)) we initialize the cloud in a strongly interacting state ($U/W=2.9(3)$) and ramp up the modulation while setting the frequency equal to the interaction $U$. 
For different driving strengths $K_0$ we measure the change of $\mathcal{D}$ for increasing ramp times \cite{si3,si4}. 
From Eq.~\eqref{Heff_res}, it is expected that the $\mathcal{D}$ creation rate scales with $J_{x} \mathcal{J}_1(K_0)$ \cite{si3}.
In the static case (green) the suppressed $\mathcal{D}$ reflects the strongly correlated regime.
In the driven system a finite density-dependent term and reduced effective interactions result in an increase of $\mathcal{D}$.
We find good agreement between theory and experiment. Both show the theoretically predicted creation of $\mathcal{D}$ scaling as $\mathcal{J}_1(K_0)$ averaged over the ramp-up in $K_0$ (see theoretical analysis in \cite{si3}).
At longer times ($t_{\text{ramp}} > 10$~ms), the renormalized tunneling and interaction energies lead to a global redistribution of density, which manifests itself in an increase of $\mathcal{D}$. 
This trap induced effect cannot be captured by nonequilibrium DMFT. 
The following decrease at $300$~ms is influenced by atom loss indicating the excitation of atoms to higher bands caused by absorption of energy from the drive \cite{si4}. 

\begin{figure}
    \includegraphics[width=\columnwidth]{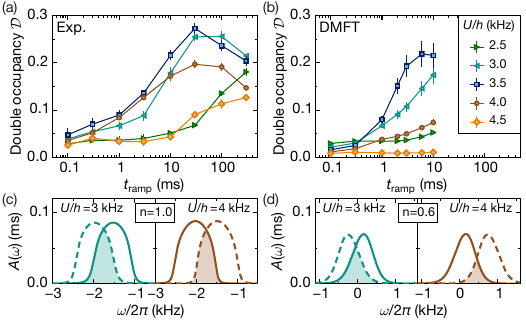}
    \caption{Near-resonant driving in the Fermi-Hubbard model for $J_{x,y,z}/h=[200(50),100(10),100(10)]$~Hz at fixed strength $K_0=1.44(2)$ and frequency $\Omega/(2\pi)=3.5$ kHz. 
    Experimental (a) and numerical results (b) for $\mathcal{D}$ after different ramp times at interactions chosen symmetrically around the resonance. Dynamics beyond $t_{\mathrm{ramp}}=10$~ms are influenced by trap effects and for $t_{\mathrm{ramp}}=$300~ms by heating \cite{si4} and is not considered in DMFT. 
    Error bars are the same as in Fig.~\ref{fig2}. 
    In (c) and (d),the occupations of the lower Hubbard band (solid lines) at $T/J_x=3.3$ are shown for symmetric detunings. 
    The shaded area indicates the overlap with the hole occupation (dashed) shifted by the driving frequency. 
    The data in (c) are for half filling and in (d) for lower filling.}\label{fig4}
\end{figure}

In another set of measurements shown in Fig.~\ref{fig4}a (experiment) and \ref{fig4}b (theory), we fix the strength ($K_0 = 1.44(2)$) and drive frequency ($\Omega/2\pi=3.5$ kHz) but change the interaction $U$ symmetrically around the resonance ($U/h=3.5$~kHz) \cite{si3,si4}. 
The far detuned data ($U/h=2.5$~kHz and $4.5$~kHz) show very low excitations of $\mathcal{D}$ for shorter ramp times, whereas in the near-resonant cases finite excitation rates appear. 
Experimentally, the curves at $U/h=3$~kHz and $U/h=4$~kHz have a comparable excitation rate to the resonant case, but a lower saturation value for $U/h=4$~kHz indicates an asymmetry of the absorption with respect to the resonance frequency.
In the DMFT data this asymmetry is already reflected in the creation rates.
At half-filling, the rate is almost symmetric, consistent with the similar size of the overlap between the occupied states and the empty states shifted by the driving frequencies $U/h=3$~kHz and $U/h=4$~kHz (Fig.~\ref{fig4}c). 
At lower fillings, since the bottom of the lower Hubbard band is more occupied, the overlap is reduced for $U/h=4$~kHz (Fig.~\ref{fig4}d), which results in the asymmetry. 
Overall, we find almost quantitative agreement between theory and experiment apart from the $U/h=4$~kHz case where the results are very sensitive to the exact Hubbard parameters which is not represented in the error bar of the theoretical calculation.
The longer ramp times ($t_{\text{ramp}} > 10$~ms), only measured in the experiment, reveal an initial increase in $\mathcal{D}$ for all detunings followed by a decrease for small detunings. This dynamics is again resulting from trap induced effects, technical heating and coupling to higher bands \cite{si4, Messer2018}.

In this work we demonstrated how basic models of nonequilibrium, strongly correlated systems can be explored experimentally and numerically to reveal their fundamental dynamics. 
New insights into pump-probe experiments in solid state physics can be gained by looking at the many-body dynamics of these strongly driven models \cite{Aoki2014,Goerg2018}. Furthermore, the cross-validation of the presented methods reveals the driving regimes where the physics is described by a desired effective Hamiltonian. In both, experiment and theory, different model Hamiltonians can be realized including a fully tunable Heisenberg and $t-J$ model \cite{Coulthard2018,Goerg2018,Mentink2015} or anyonic Hubbard models and dynamical gauge fields resulting from occupation dependent Peierls phases \cite{Keilmann2011,Greschner2015,Cardarelli2016,Strater2016a,Greschner2014,Bermudez2015,Barbiero2018}.

\vspace{1mm}

\begin{acknowledgments}
We thank W. Zwerger for encouraging this collaboration, and J. Coulthard, H. Gao, D. Gole\v{z}, D. Jaksch, M. Sch\"{u}ler, and K. Viebahn for helpful discussions. We acknowledge the Swiss National Science Foundation (Project Number 169320 and 182650, NCCR-QSIT and NCCR-MARVEL), the Swiss State Secretary for Education, Research and Innovation Contract No. 15.0019 (QUIC), ERC advanced grant TransQ (Project Number 742579) and ERC consolidator grant Modmat (Project Number 724103) for funding. The DMFT calculations have been performed on the Beo04 and Beo05 clusters at the University of Fribourg. 
\end{acknowledgments}

\bibliography{references}

\clearpage

\makeatletter
\setcounter{section}{0}
\setcounter{subsection}{0}
\setcounter{figure}{0}
\setcounter{equation}{0}
\renewcommand{\bibnumfmt}[1]{[S#1]}
\renewcommand{\thefigure}{S\@arabic\c@figure}
\renewcommand{\theequation}{S\@arabic\c@equation}
\makeatother

\renewcommand{\theHfigure}{S\thefigure}
\renewcommand{\theHequation}{S\theequation}

\makeatother

\centerline{\Large \textbf{Supplemental material}}

\section{Nonequilibrium DMFT simulations}
\label{NDMFT}
\subsection{Model}
\label{NDMFT_Model}
\begin{figure}[t]
 \centering
   \hspace{-0.cm}
    \vspace{0.0cm}
   \includegraphics[width=50mm]{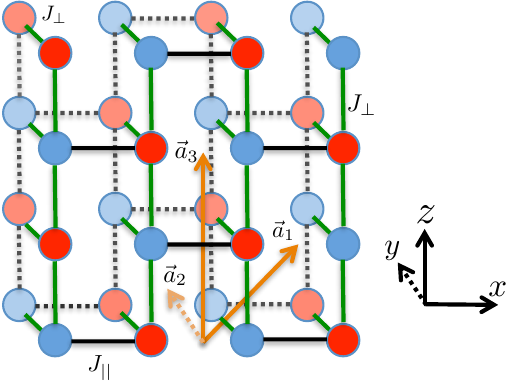} 
  \caption{Schematic picture of the three-dimensional hexagonal lattice. Red and blue circles correspond to A, B sublattices, respectively. }
  \label{fig:lattice_sche}
\end{figure}

We consider the three-dimensional brick wall lattice illustrated in Fig.~\ref{fig:lattice_sche}, which is characterized by the primitive vectors (${\bf a}_1,{\bf a}_2$ and ${\bf a}_3$).
The unit cell consists of two sites (A, B sublattices) and in the following we denote the position of a site in cell $i$ and sublattice $\alpha$ by ${\bf r}_{i,\alpha}$. 
The position of a cell is specified by ${\bf r}_{i}={\bf r}_{i,\alpha}-{\bf r}_{0,\alpha}$, which can be 
expressed as ${\bf r}_{i}=i_1 {\bf a}_1 + i_2 {\bf a}_2 + i_3 {\bf a}_3$, where $i_w\in {\bf Z}$. 
The corresponding primitive vectors of the reciprocal lattice are introduced as usual, for example,
${\bf b}_1=2\pi \frac{{\bf a}_2 \times {\bf a}_3}{{\bf a}_1\cdot ({\bf a}_2\times {\bf a}_3)}$.
Considering a lattice system consisting of $L_w$ unit cells for each ${\bf a}_w$-direction
with periodic boundary conditions in each direction,
we can introduce discretized momentum vectors in the reciprocal space as 
\begin{align}
{\bf k}_l=\frac{l_1}{L_1}{\bf b}_1+\frac{l_2}{L_2}{\bf b}_2+\frac{l_3}{L_3}{\bf b}_3, 
\end{align}
with $l=(l_1,l_2,l_3)$ and $l_w=0,1,\cdots L_w-1$, such that ${\bf k}_l\cdot{\bf r}_i=2\pi \sum_{w}\frac{i_w l_w}{L_w}$.
The electron creation operators in momentum space are defined as 
\begin{align}
\hat{c}^\dagger_{{\bf k}_l,\alpha,\sigma}=\frac{1}{\sqrt{N}} \sum_i e^{i {\bf k}_l\cdot{\bf r}_i} \hat{c}^\dagger_{{\bf r}_i,\alpha,\sigma},
\end{align}
where $N=L_1L_2L_3$ and $\sigma$ the spin. With these, the kinetic part of the Hamiltonian for this lattice can be written as 
\begin{widetext}
\begin{subequations}\label{eq:3dhex}
\begin{align}
\hat{H}_0(t)&=\sum_{{\bf k},\sigma}
\begin{bmatrix}
\hat{c}^\dagger_{{\bf k},A,\sigma} & \hat{c}^\dagger_{{\bf k},B,\sigma}
\end{bmatrix}
\hat{h}_{\bf k}(t)
\begin{bmatrix}
\hat{c}_{{\bf k},A,\sigma} \\ \hat{c}_{{\bf k},B,\sigma}
\end{bmatrix},\\
\small
\hat{h}_{\bf k}(t)&=
\begin{bmatrix}
 -2J_{\perp} \cos({\bf k}\cdot {\bf a}_3) & -J_{||}^*(t)-J_{\perp}(e^{i{\bf k}\cdot {\bf a}_2}+1)e^{-i{\bf k}\cdot {\bf a}_1}\\
-J_{||}(t)-J_{\perp}(e^{-i{\bf k}\cdot {\bf a}_2}+1)e^{i{\bf k}\cdot {\bf a}_1} & -2J_{\perp} \cos({\bf k}\cdot {\bf a}_3)
\end{bmatrix}.
\normalsize
\end{align}
\end{subequations}
\end{widetext}
Here $J_{||}=J_x$ and $J_{\perp}=J_y=J_z$ in the main text.
The effect of the external field along the $x$-axis is taken into account through the Peierls substitution 
\begin{align}
-J_{||}(t)=-J_{||}\exp\Big[-iq {\bf A}(t)\cdot({\bf a_1}-\tfrac12{\bf a_2}) \Big].\label{eq:peierls}
\end{align}
The vector potential ${\bf A}(t)$ is related to the electric field by ${\bf E}(t)=-\hbar\partial_t {\bf A}(t)$.
In the present study we choose $q=1$ and $|{\bf a_1}-\frac{1}{2}{\bf a_2}|=d_x$ and apply the field along ${\bf a_1}-\frac{1}{2}{\bf a_2}$ $(\equiv {\bf e}_x)$.
The strength and time-dependence of the field are the same as in the experiment,
\begin{subequations}\label{eq:ramp}
\begin{align}
E(t)&=\frac{K_0}{d_x}\cdot \hbar\Omega\;F_{\rm ramp}(t;t_{\rm ramp})\sin(\Omega t),\\
F_{\rm ramp}(t;t_{\rm ramp})&=
\begin{cases}
t/t_\text{ramp} \;\;\;\;\;\;\; t\in[0,t_\text{ramp}] \\
1 \;\;\;\;\;\;\;\;\;\;\; t>t_{\rm ramp}
\end{cases}.
\end{align}
\end{subequations}

Before we proceed, we introduce a useful expression of the kinetic term.
Assuming that the ramp is over or slow enough that the change of the field amplitude is negligible on the timescale of the oscillations, we can express the vector potential as ${\bf A}(t)=K_0 \cos(\Omega t) {\bf e}_x$.
Expanding Eq.~\eqref{eq:peierls} into harmonics one obtains 
\begin{align}\label{eq:H0_express2}
&\hat{H}_0(t)=-J_{\perp}\sum_{\langle {\bf i},{\bf j}\rangle_{y,z},\sigma} \hat{c}^\dagger_{{\bf i},\sigma} \hat{c}_{{\bf j},\sigma}
-J_{||}\mathcal{J}_0(K_0) \sum_{\langle {\bf i},{\bf j}\rangle_x,\sigma}   \hat{c}^\dagger_{{\bf i},\sigma} \hat{c}_{{\bf j},\sigma}\nonumber\\
&-J_{||}\sum_{\langle {\bf i},{\bf j}\rangle_x,\sigma} \sum_{n\in{{\bf Z}}, n\neq 0} e^{in\frac{\pi}{2}} \mathcal{J}_n(K_0 {\bf e}_x\cdot {\bf r}_{{\bf ij}}) e^{in\Omega t} \hat{c}^\dagger_{{\bf i},\sigma} \hat{c}_{{\bf j},\sigma}.
\end{align}
Here ${\bf i}$ and ${\bf j}$ indicate sites regardless of the sublattice, i.e. they denote cell and sublattice indices.

The total Hamiltonian consists of $\hat{H}_0(t)$, the chemical potential term and the interaction term: 
$\hat{H}_{\rm tot}(t)=\hat{H}_0(t)-\mu \hat{N}+H_{\rm int}$.
In our case, the interaction term is the local Hubbard term, Eq.~(1) in the main text.
In the theoretical study, we use $J_x/h=200$~Hz as the unit of energy and choose $J_y,\;J_z,\;U$ and $\Omega$ according to the experiment.
The systems considered in this study are in the strongly-correlated regime.
We solve the model with single-site DMFT, using a perturbative strong-coupling impurity solver (see the next section).
It turns out that $(L_1,L_2,L_3)$=$(6,6,6)$ is large enough to reach convergence in the systems size.

\subsection{Single-site DMFT}
\label{NDMFT_Single}
In order to simulate the time-evolution of the Fermi-Hubbard model on the three-dimensional brick wall lattice
we employ single-site DMFT.
For this, we define the Green's functions in real space and momentum space as
\begin{subequations}
\begin{align}
G_{ij,\alpha\beta}(t,t')&=-i\langle T_{\mathcal{C}} \hat{c}_{{\bf r}_i,\alpha,\sigma}(t) \hat{c}^\dagger_{{\bf r}_j,\beta,\sigma}(t')\rangle,\\
\hat{G}_{\bf k}(t,t')&=-i
\left\langle T_{\mathcal{C}} 
\begin{bmatrix}
\hat{c}_{{\bf k},A,\sigma}(t) \\ \hat{c}_{{\bf k},B,\sigma}(t)
\end{bmatrix}
\begin{bmatrix}
\hat{c}^\dagger_{{\bf k},A,\sigma} (t')& \hat{c}^\dagger_{{\bf k},B,\sigma}(t')
\end{bmatrix}
\right \rangle,
\end{align}
\end{subequations}
where $\mathcal{C}$ indicates the Kadanoff-Baym (KB) contour and $T_{\mathcal{C}}$ is the contour ordering operator \cite{Aoki2014}.
Here we omitted the spin index on the left hand side assuming spin symmetry of the system.
Within single-site DMFT, we assume a local self-energy on each sublattice,
\begin{align}
\Sigma_{ij,\alpha\beta}(t,t')=\delta_{i,j}\delta_{\alpha,\beta} \Sigma_{\alpha}(t,t').
\end{align}
Under this assumption, the Green's functions in momentum space can be calculated as 
\begin{align}
&\hat{G}_{\bf k}(t,t')=[(i\hbar \partial_t\hat{I}+\mu\hat{I}-\hat{h}_{\bf k}(t))\delta_{\mathcal{C}}(t,t')-\hat{\Sigma}(t,t')]^{-1},
\end{align}
with $\delta_C$ the delta-function on the KB contour,
\begin{align}
\hat{\Sigma}(t,t')=
\begin{bmatrix}
\Sigma_{A}(t,t')&0\\
0& \Sigma_{B}(t,t')
\end{bmatrix},\nonumber
\end{align}
and the local Green's functions is 
\begin{align}
G_{ii,\alpha\alpha}(t,t')=\frac{1}{N} \sum_{\bf k} G_{{\bf k},\alpha \alpha}(t,t').\label{eq:imp_dyson1}
\end{align}

In DMFT, the self-energies $\Sigma_\alpha$ are evaluated by introducing an effective impurity problem for each sublattice,
\begin{subequations}
\begin{align}
S_{{\rm imp},\alpha}&=i\int_{\mathcal{C}} dt dt' \sum_\sigma d^\dagger_{\sigma}(t) \mathcal{G}_{\alpha}^{-1}(t,t') d_\sigma(t')\nonumber\\
&\;\;\;\;-i\int_{\mathcal{C}} dt H_{\rm int}(t),\\
\mathcal{G}_{\alpha}^{-1}(t,t')&=(i\hbar \partial_t+\mu)\delta_{\mathcal{C}}(t,t')-\Delta_\alpha (t,t').
\end{align}
\end{subequations}
Here $S_{\rm imp}$ is the action of the impurity problem in the path integral formalism and $d^\dagger$ indicates the Grassmann number 
associated with the creation operator of the electron on the impurity site. The noninteracting impurity Green's functions $\mathcal{G}_{\alpha}$, or equivalently the hybridization functions $\Delta_\alpha$, play the role of dynamical mean fields. The interacting (spin-independent) impurity Green's function is related to the impurity self-energy $\Sigma_{\rm imp,\alpha}$ by 
\begin{align}
G_{\rm imp,\alpha}(t,t')=[\mathcal{G}_{\alpha}^{-1}(t,t')-\Sigma_{\rm imp,\alpha}(t,t')]^{-1}.
\end{align}
In the DMFT self-consistency loop, the dynamical mean fields are determined  such that  
$\Sigma_{\rm imp,\alpha}=\Sigma_{\alpha}$ and 
\begin{align}
G_{\rm imp,\alpha}(t,t')=G_{ii,\alpha\alpha}(t,t').
\end{align}
Once convergence is reached we thus obtain the lattice self-energy by calculating the impurity self-energy. 

\begin{figure}[t]
 \centering
   \hspace{-0.cm}
    \vspace{0.0cm}
   \includegraphics[width=85mm]{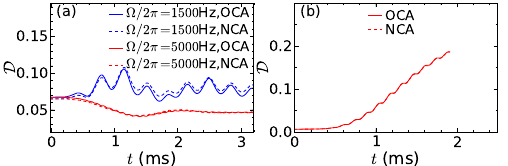} 
  \caption{Comparison of $\mathcal{D}$ obtained using NCA and OCA in the off-resonant regime (a) and the near-resonant regime (b). 
  For the off-resonant regime, we use $J_x/h=200$~Hz, $J_y/h=J_z/h=40$~Hz, $U/h=750$~Hz, $K_0=1.43$ and $t_{\rm ramp}=1.0$~ms at $T/J_x=1.21$.  
  For the near-resonant regime, we use $J_x/h=200$~Hz, $J_y/h=J_z/h=100$~Hz, $U/h=\Omega/(2\pi)=3500$~Hz, $K_0=1.43$, and $t_{\rm ramp}=1.0$~ms at $T/J_x=3.3$.  }
  \label{fig:fig_oca_nca}
\end{figure}

To solve the impurity problem, we mainly employ the non-crossing approximation (NCA), which is the lowest order expansion in the hybridization function and should be good in the strong coupling regime \cite{Eckstein2010}. It is known that this approximation starts to deviate from numerically exact results in the intermediate-coupling regime at low temperatures, where the one-order higher expansion (one-crossing approximation, OCA) provides important corrections \cite{Eckstein2010} at the cost of a much higher computational expense. In Fig.~\ref{fig:fig_oca_nca}, we compare the results for $\mathcal{D}$ from NCA and OCA implementations for some representative off-resonant and near-resonant conditions.  The results show that for the interactions and temperatures considered in this study, NCA and OCA yield quantitatively almost the same results (the difference is at most 5\%), which justifies the use of NCA in the analysis.

\subsection{Observables}
\label{NDMFT_observables}
The density of particles is calculated as 
\begin{align}
n({\bf r}_i,t)=\frac{1}{2} \sum_{\alpha=A,B}\sum_{\sigma=\uparrow,\downarrow}
\langle \hat{n}_{{\bf r}_{i,\alpha},\sigma}(t)\rangle.
\end{align}
In the present study, we focus on the double occupation,
\begin{align}
n_d({\bf r}_i,t)=\frac{1}{2} \sum_{\alpha=A,B}\langle \hat{n}_{{\bf r}_{i,\alpha},\uparrow}(t) \hat{n}_{{\bf r}_{i,\alpha},\downarrow}(t)\rangle,
\end{align}
which is averaged over the sublattices.
In each DMFT simulation, a homogenous system is considered, hence the DMFT results show no dependence on ${\bf r}_i$.
By running simulations for different chemical potentials at the experimentally estimated temperature (\ref{Exp_density}), we obtain the double occupation as a function of the density of particles, which 
we express as $n_d[n;t]$.

In the experiment, the system is inhomogeneous because of the trapping potential, and 
the double-occupation ($\mathcal{D}$) is measured as the number of doubly occupied sites normalized by the total number of particles.
In order to capture the effect of this inhomogeneous set-up, we employ the local density approximation.
Namely, we average the results of DMFT simulations for homogeneous systems using the experimental density profile: 
\begin{align}
\mathcal{D}(t)&=\frac{\int 2n_d[n({\bf r});t] dxdydz}{\int n({\bf r}) dxdydz},
%\mathcal{D}&=\frac{\int 2n_d({\bf r}) dxdydz}{\int n({\bf r}) dxdydz},
%&=\frac{\int 2n_d[n({\bf r})] r^2\sin\theta drd\phi d\theta}{\int n({\bf r}) r^2\sin\theta drd\phi d\theta},
\end{align}
where ${\bf r}=(x,y,z)$.
This procedure is justified on timescales which are short enough that the redistribution of atoms in the trap can be neglected, which is the case here. 

In DMFT, one can evaluate the equilibrium local spectral function
\begin{align}
A(\omega)\equiv -\frac{1}{2\pi}\sum_{\alpha=A,B} {\rm Im} \int dt_\text{rel} G_{ii,\alpha\alpha}^R(t_\text{rel})e^{i\omega t_\text{rel}}
\end{align}
from the Fourier transform of the equilibrium retarded Green's functions. 
Here $t_{\rm rel}=t-t'$ is the relative time of the two time arguments in the Green's functions.
The particle and hole occupation are
\begin{align}
N(\omega)=A(\omega)f(\omega),\;\;
\bar{N}(\omega)&=A(\omega)\bar{f}(\omega),
\end{align}
where $f(\omega)=\frac{1}{1+e^{\beta\hbar\omega}}$ is the Fermi distribution function and $\bar{f}(\omega)=1-f(\omega)$ for inverse temperature $\beta$.

In the nonequilibrium case, we measure the local spectral function and the occupation as 
\begin{align}
A(\omega;t_{\rm av})\equiv -\frac{1}{2\pi}\sum_{\alpha=A,B} {\rm Im} \int dt_\text{rel} 
G_{ii,\alpha\alpha}^R(t_{\rm rel}; t_{\rm av})e^{i\omega t_\text{rel}},
\end{align}

\begin{align}
N(\omega;t_{\rm av})\equiv \frac{1}{4\pi}\sum_{\alpha=A,B} {\rm Im} \int dt_\text{rel} 
G_{ii,\alpha\alpha}^<(t_{\rm rel}; t_{\rm av})e^{i\omega t_\text{rel}}.
\end{align}
Here $<$ indicates the lesser part of the Green's function, $t_{\rm av}=\frac{t+t'}{2}$ is the average time of the two time arguments in the Green's functions and $G(t_{\rm rel}; t_{\rm av})= G(t,t')$.
The hole occupation is defined as $\bar{N}(\omega;t_{\rm av}) = A(\omega;t_{\rm av})-N(\omega;t_{\rm av})$.

We also introduce the current operator along the $x$ direction and its current-current response function, 
\begin{align}
&\hat{J}_x=qJ_{||}\sum_{{\bf k},\sigma} 
\begin{bmatrix}
\hat{c}^\dagger_{{\bf k},A,\sigma} (t')& \hat{c}^\dagger_{{\bf k},B,\sigma}
\end{bmatrix}
\hat{\sigma}_2
\begin{bmatrix}
\hat{c}_{{\bf k},A,\sigma} \\ \hat{c}_{{\bf k},B,\sigma}
\end{bmatrix},\\
&\chi_{xx}(t,t')=-i\langle T_{\mathcal{C}} \hat{J}_x(t) \hat{J}_x(t') \rangle.
\end{align}
Here $\hat{\sigma}_2$ is the Pauli matrix.
The retarded part of $\chi_{xx}(t,t')$ is the response function, and the imaginary part of $\chi^R_{xx}(\omega)$ corresponds 
to the probability of energy absorption from weak excitation fields. 
If we neglect the vertex correction (this is exact in simple lattices such as cubic and hyper-cubic lattices), $\chi_{xx}(t,t')$ can be expressed with the single particle Green's functions as
\begin{align}
\chi_{xx}(t,t')=-iq^2J_{||}^2 \sum_{\bf k}\text{tr}[\hat{\sigma}_2 \hat{G}_{\bf k}(t,t') \hat{\sigma}_2 \hat{G}_{\bf k}(t',t)].
\end{align}
This leads to
\begin{align}
-\frac{1}{\pi}{\rm Im}\chi_{xx}&(\Omega)=
q^2J_{||}^2 \sum_{\bf k} \int d\omega {\rm tr}[\hat{\sigma}_2 \hat{A}_{\bf k}(\omega+\Omega) \hat{\sigma}_2 \hat{A}_{\bf k}(\omega)]\nonumber\\
& \;\times[f(\omega)\bar{f}(\omega+\Omega)-\bar{f}(\omega)f(\omega+\Omega)].\label{eq:chi_w}
\end{align}
Here $\hat{A}_{\bf k}(\omega)\equiv -\frac{1}{2\pi}{\rm Im}[\hat{G}^R_{\bf k}(\omega)-\hat{G}^A_{\bf k}(\omega)]$, which is the single particle spectrum, and $R$ and $A$ indicate the retarded and advanced parts.
The first term corresponds to absorption and the second term to emission. 
In the present study, due to $\hbar\Omega\beta\gg 1$ we can neglect the emission term.
In addition, in the strong coupling regime, the support of $A_{\bf k}(\omega)$ turns out to be almost the same as $A(\omega)$,
so that it is enough to consider $A(\omega)$ to roughly estimate the possible absorption process.
\section{Off-resonant driving}
\label{NDMFT_OR}
In this part, we choose $J_x=1,J_y=J_z=0.2$ and $U=3.75$, which correspond to $200$~Hz, $40$~Hz and $750$~Hz, respectively. We use the inverse temperature of $\beta=0.83$, which is estimated by the experiment in a preparation lattice (\ref{Exp_density}). In addition to the initial lattice configuration stated above, $\mathcal{D}$ is also measured for a static reference lattice with a reduced tunneling $J_x=0.4$ (corresponds to $80$~Hz) in the experiment. Since the interaction is the same for both configurations, $U/W$ increases from $1.1$ for $J_x=1$ to $1.6$ for $J_x=0.4$. With the larger value of $U/W$, the temperature estimation at equal entropy is more difficult, hence, direct evaluation of the reference value from DMFT becomes also difficult. Thus, the best estimation for $\mathcal{D}$ in local equilibrium is provided by an adiabatic ramp from $J_x=1$ to $J_x=0.4$. We use a ramp time of $5$~ms where saturation at $\mathcal{D}=0.043$ is reached (see Fig.~2).

In the lowest order $1/\Omega$-expansion, the effective static Hamiltonian (after the ramp) is obtained by neglecting the oscillating terms 
in Eq.~\eqref{eq:H0_express2}, where the hopping parameter of particles along the $x$ direction is renormalized by $\mathcal{J}_0(K_0)$.
The bandwidth of the free particle system is thus reduced from $W=2J_{||}+4J_{\perp}$ to $W'=2\mathcal{J}_0(K_0)J_{||}+4J_{\perp}$.
The neglected terms can be regarded as additional excitation terms of this effective static Hamiltonian.

In the experiment, the external field is introduced with finite ramp times, Eq.~\eqref{eq:ramp}.
If the ramp speed is slow enough, it is expected that the dynamics can still be described by an effective 
Hamiltonian with a time-dependent hopping in the $x$ direction but without time-periodic excitation,
\begin{align}
&-J_{||}\mathcal{J}_0(K_0) \sum_{\langle {\bf i},{\bf j}\rangle_x,\sigma}   \hat{c}^\dagger_{{\bf i},\sigma} \hat{c}_{{\bf j},\sigma} \nonumber\\
&\rightarrow
-J_{||}\mathcal{J}_0(K_0\cdot F_{\rm ramp}(t;t_{\rm ramp}) ) \sum_{\langle {\bf i}, {\bf j}\rangle_x,\sigma}   \hat{c}^\dagger_{{\bf i},\sigma} \hat{c}_{{\bf j},\sigma}.
\end{align}
We call this the ``protocol 1" excitation.
On the other hand, in the experiments, a linear ramp of the hopping parameter in the $x$ direction is implemented:
\begin{align}
&-J_{||}\mathcal{J}_0(K_0) \sum_{\langle {\bf i},{\bf j}\rangle_x,\sigma}   \hat{c}^\dagger_{{\bf i},\sigma} \hat{c}_{{\bf j},\sigma} \\
&\rightarrow
-J_{||}(1+(\mathcal{J}_0(K_0)-1)\cdot F_{\rm ramp}(t;t_{\rm ramp})) \sum_{\langle {\bf i},{\bf j}\rangle_x,\sigma}   \hat{c}^\dagger_{{\bf i},\sigma} \hat{c}_{{\bf j},\sigma}\nonumber.
\end{align}
We call this the ``protocol 2" excitation. 
In the following, we compare the results of the periodic excitation and these two excitation protocols.

\subsection{Real time propagation}
\label{NDMFT_OR_propagation}

\begin{figure}[t]
 \centering
   \hspace{-0.cm}
    \vspace{0.0cm}
	\includegraphics[width=85mm]{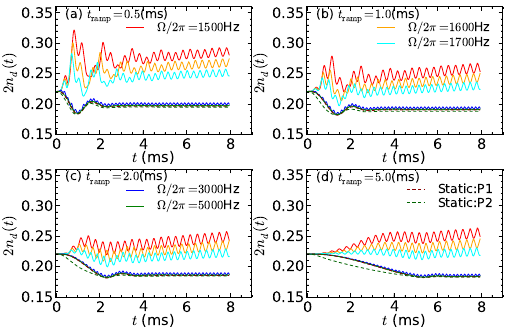}
  \caption{Time evolution of the double occupation $2n_d(t)$ for different excitation protocols and ramp times.
  Solid lines correspond to the time-periodic excitations with specified frequencies, while the dashed lines correspond to the modulations
 of the hopping parameter (see the text).
  Here, $\beta=0.83$, $K_0=1.43$ and $n=1.0$. P1 and P2 stands for protocol 1 and protocol 2.}
  \label{fig:ACI_Heff_compare_b075n1}
\end{figure}

In this section, we present the time-evolution of the double occupation ($2n_d(t)$) simulated by DMFT with a fixed density.
In Fig.~\ref{fig:ACI_Heff_compare_b075n1}, we show the results for different excitation protocols and ramp times at half-filling.
For all ramp times studied here, the results of $\Omega/(2\pi)=3000$~Hz and $\Omega/(2\pi)=5000$~Hz almost match those of the hopping modulation protocol 1,
with $\Omega/(2\pi)=5000$~Hz showing a better agreement, as expected.
In particular, it is interesting to point out that the description based on the static Floquet Hamiltonian, time-dependent effective hopping parameter is meaningful even when the amplitude of the excitation is modulated 
rather quickly compared to the excitation frequency. 
For example, for $t_{\rm ramp}=1.0$ ms, there are only three oscillations for  $\Omega/(2\pi)=3000$~Hz and five oscillations for $5000$~Hz during the ramp of the field 
strength. Nevertheless, the dynamics predicted by the Floquet Hamiltonian reproduces the full dynamics rather well. 
The results of the hopping modulation protocol 2 show a somewhat different dynamics during the ramp, but the values of the double occupation reached at the end of ramp and 
the dynamics after the ramp are very similar to the results obtained using the protocol 1. This tendency is particularly clear as we increase the ramp time.

\begin{figure}[t]
 \centering
   \hspace{-0.cm}
    \vspace{0.0cm}
   \includegraphics[width=65mm]{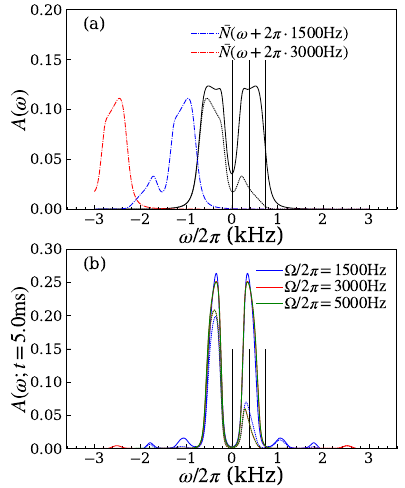} 
  \caption{(a) Local single-particle spectral function ($A(\omega)$, black curve) and its occupation ($N(\omega)$, black dashed curve) at $\beta=0.83$ in equilibrium at half-filling. The unoccupied parts $\bar N$ of the spectral function shifted by $1500$~Hz and $3000$~Hz are also shown to illustrate possible absorption processes in the driven state (dash-dotted lines). (b) Nonequilibrium local single-particle spectral function measured at $t_{\rm av}=5.0$~ms for $K_0=1.43$, $t_{\rm ramp}=5.0$~ms, $n=1$ and  different excitation frequencies $\Omega$. The vertical lines indicate $\frac{U}{2}-\frac{W}{2},\frac{U}{2}$ and $\frac{U}{2}+\frac{W}{2}$.}
  \label{fig:Akw_U375b075_excitation}
\end{figure}

Now we reduce the excitation frequency down to $1500$~Hz, which is just above the naively expected bandwidth $(U+W)/h=1470$~Hz
and that of the Floquet Hamiltonian after the ramp $(U+W')/h=1230$~Hz.
We clearly see large deviations from the prediction of the Floquet Hamiltonian, and in particular a double occupation which  continuously increases.
There are several points to note here.
Firstly, around this frequency range the nonequilibrium dynamics strongly depends on the excitation frequency, compare the results for $1500,1600,1700$~Hz.
Secondly, the production of double occupations is particularly pronounced during the ramp, while after the ramp the increase becomes more moderate.
Thirdly, the shorter the ramp time, the larger the increase of the double occupation during the ramp.
The last point manifests itself as a larger value of $2 n_d$ averaged over one period after the ramp in the experiment.
The first and second points can be related to the local spectral function, and we come back to this point in the next section.
The third observation is explained by the fact that for the shorter ramp, the field includes higher frequency components than $\Omega$ during the ramp, which can lead to efficient absorption of energy from the field. 
We also note that the results away from half-filling ($n=0.6$) show the same qualitative behavior, including the same characteristic time scales.

\subsection{Analysis of the single-particle spectrum}
\label{NDMFT_OR_spectrum}
Here, we discuss the properties of the single-particle spectrum of the many-body state, which is difficult to access experimentally.
In Fig.~\ref{fig:Akw_U375b075_excitation}(a), we show the DMFT result for the local spectrum ($A(\omega)$) and its occupation in equilibrium at half filling (black curves). 
Compared to the naive expectation that the total bandwidth is just $U+W$, it is larger because of correlation-induced broadening.
In the figure, we also show the spectral functions shifted by $1500$~Hz and $3000$~Hz.
As explained above, the overlap between the shifted unoccupied part of the spectrum and the occupation can be roughly connected to the efficiency of the absorption under a weak excitation with frequency corresponding to the shift.
For $1500$~Hz, there is a non-negligible overlap because of the broadened spectral function, while for $3000$~Hz there is almost no overlap.
These results are consistent with the doublon dynamics observed in the previous section.
In particular, when the excitation frequency is comparable to the width of the spectrum, a small change of the 
frequency substantially changes the overlap area, which leads to a sensitive dependence of the dynamics on the excitation frequency.

Moreover, when the excitation amplitude
becomes stronger, the bandwidth of the upper and lower bands is reduced, which leads to a reduction of the total bandwidth and hence the strength of the absorption. 
To illustrate this effect, we plot in Fig.~\ref{fig:Akw_U375b075_excitation}(b) the local spectral function $A(\omega;t)$,  measured after completion of the ramp, which indeed exhibits a substantial reduction of the bandwidth. This can explain the more moderate increase of the double occupation after the ramp.

\section{Near-resonant driving}
\label{NDMFT_NR}
Here we consider the case where $U \approx \hbar\Omega$, where the efficient creation of $\mathcal{D}$ is expected \cite{Herrmann2018, Peronaci2018}.
In this case, we can introduce the effective static Hamiltonian Eq.~(3) in the rotating frame \cite{Bukov2016b}.
In this section, we choose $J_x=1,J_y=J_z=0.5$ and $U=17.5$, which correspond to $200$~Hz, $100$~Hz and $3500$~Hz, respectively. We use the inverse temperature of $\beta=0.30$, which is estimated by the experiment (\ref{Exp_density}).
\begin{figure}[t]
 \centering
   \hspace{-0.cm}
    \vspace{0.0cm}
	\includegraphics[width=85mm]{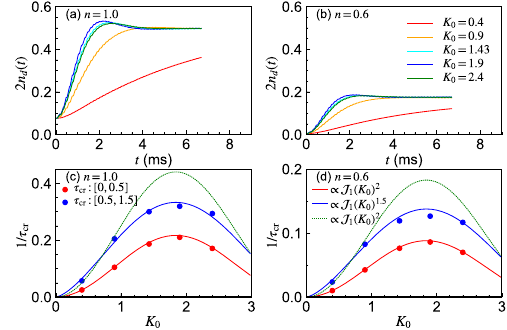}
  \caption{
    (a)(b) Time evolution of the double occupation $2n_d(t)$ after the AC quench with different field strengths. 
  (c)(d) Doublon creation rates estimated from the slopes of the linear fits for the indicated time intervals and for different field strengths.
  We also show lines proportional to $\mathcal{J}_1(K_0)^2$ and $\mathcal{J}_1(K_0)^{1.5}$.
  Panel (a)(c) is for $n=1.0$ and panel (b)(d) is for $n=0.6$. Here, $U/h=\Omega/(2\pi)=3500$~Hz and $\beta=0.30$.}
  \label{fig:AC_QUENCH_U175b03_tot}
\end{figure}

\subsection{Quench dynamics: Doublon creation ratio}
\label{NDMFT_NR_quench}
Before we show the time evolution under the excitation with finite ramp time, 
we analyze the dynamics after an AC quench for $U=\hbar\Omega$, i.e. $t_{\rm ramp}=0$.
In this case, the effective Hamiltonian (in the rotating frame) exhibits no time dependence,
and it implies that the hopping parameter relevant for the doublon-holon creation is proportional to $\mathcal{J}_1(K_0)$.
One can reach the same conclusion from Eq.~(\ref{eq:H0_express2}) by regarding the $n=\pm1$ terms as perturbations and neglecting higher harmonic terms.
Since $\mathcal{J}_1(K_0)$ exhibits a maximum around $K_0\simeq 1.85$, the doublon creation rate 
should exhibit a maximum around this value.
Indeed, we can see this in the DMFT analysis for different fillings, see Fig.~\ref{fig:AC_QUENCH_U175b03_tot}(a)(b).
The fastest increase of the doublon number is observed for $K_0=1.9$ regardless of the density.
Initially, the increase shows a super-linear (quadratic) behavior for all cases considered here.
This behavior can be understood by considering the time evolution from the state  
$\hat{c}^\dagger_{1,\uparrow}\hat{c}^\dagger_{2,\downarrow}|{\rm vac}\rangle$ in a dimer system. The time evolution in this system results in an increase of the doubly occupied states $\propto \sin^2(2J_x \mathcal{J}_1(K_0)t/\hbar)$.

We also extracted the doublon creation rate by linear fitting, $n_d(t)=t/\tau_{\rm cr} + a$.
In Fig.~\ref{fig:AC_QUENCH_U175b03_tot}(c)(d), we show $1/\tau_{\rm cr}$ extracted using two different fitting ranges.
One can see that at the very early stage, $1/\tau_{\rm cr}$ scales as $\mathcal{J}^2_1(K_0)$, as is expected from 
the Femi golden rule.
On the other hand, $1/\tau_{\rm cr}$ deviates from $\mathcal{J}^2_1(K_0)$  and better matches $\mathcal{J}^{1.5}_1(K_0)$
at later times. 

\begin{figure}[t]
 \centering
   \hspace{-0.cm}
    \vspace{0.0cm}
	\includegraphics[width=85mm]{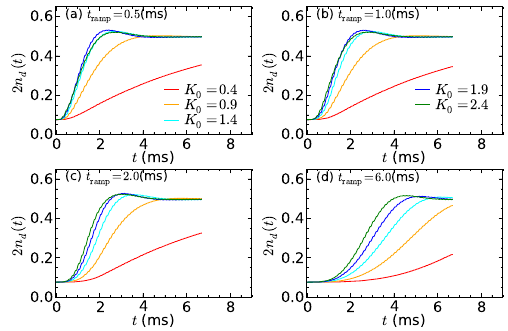}
  \caption{Time evolution of the double occupation $2n_d(t)$ for different field strengths and ramp times.
  Here, $U/h=\Omega/(2\pi)=3500$~Hz, $\beta=0.30$ and $n=1.0$.
  }
  \label{fig:ACI_Heff_compare_U175b1n1}
\end{figure}

\subsection{Real-time propagation}
\label{NDMFT_NR_propagation}
Now we move on to the results for the finite ramp time, which are relevant for the comparison to the experiments.
In Fig.~\ref{fig:ACI_Heff_compare_U175b1n1}, we show the time evolution of the double occupation $2n_d(t)$ for different field strengths and ramp times at half filling.
When $t_{\rm ramp}$ is small, one can see the fastest increase at $K_0=1.9$ as in the quench case, see $t_{\rm ramp}=0.5$~ms.
As the ramp time increases, $K_0=2.4$ starts to show a faster increase.
This is because for larger $K_0$ the system experiences a ``decent field strength" for a longer period of time during the ramp. 
In practice, $K_0\in[1.4,2.4]$ results in a fast increase of similar magnitude.
If we compare the ramp of the field up to $K_0=1.9$ to the ramp up to $K_0=2.4$ for the same ramp time,
the latter includes a longer time period where $K_0$ is in $[1.4,2.4]$. 
The results for the doped system ($n=0.6$) show the qualitatively same behavior, including the same characteristic time scales. 

\begin{figure}[t]
 \centering
   \hspace{-0.cm}
    \vspace{0.0cm}
   	\includegraphics[width=85mm]{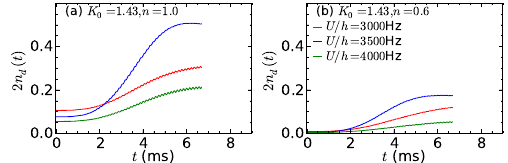}
  \caption{Time evolution of the double occupation $2n_d(t)$ for different $U$ around $\Omega/(2\pi)=3500$~Hz. Here, we choose $t_{\rm ramp}=6.0$~ms, $K_0=1.43$, and $\beta=0.30$.}
  \label{fig:AC_EXP_RAMP_U175b1_tr754_detune2_tmp}
\end{figure}

In Fig.~\ref{fig:AC_EXP_RAMP_U175b1_tr754_detune2_tmp}, we show the effects of detuning around $U\simeq \hbar \Omega$.
The fastest increase of the double occupation can be observed at $U=\hbar\Omega$.
At half-filling, the sign of the detuning $U-\hbar\Omega$ has some influence on the doublon production, but the difference between the $U-\hbar\Omega>0$ and $U-\hbar\Omega<0$ regimes is not so large.
We point out that a similar dependence of the doublon creation was observed near the other resonant condition $U\simeq 2\hbar\Omega$ in \cite{Herrmann2018},
where the difference between the two regimes was shown to becomes negligible as $U$ increases. 
On the other hand, away from half-filling, the asymmetry between the positive and negative detuning results becomes prominent, while the difference between $U/h=3000$~Hz and $U/h=3500$~Hz becomes less prominent,
which can be explained by analyzing the properties of the spectral functions, as discussed in the next section.
We also note that the difference between positive and negative detuning becomes more prominent for lower temperatures.

\begin{figure}[t]
 \centering
   \hspace{-0.cm}
    \vspace{0.0cm} 
   	\includegraphics[width=85mm]{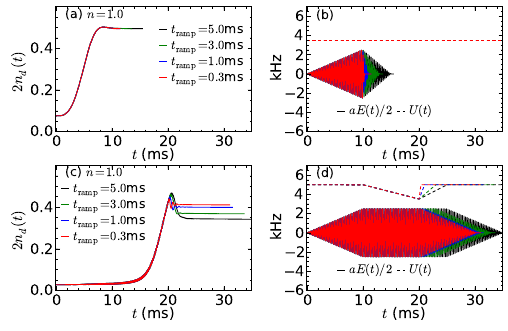}
  \caption{
  Time evolution for different adiabatic protocols at half-filling.
  (a) Double occupation $2n_d(t)$ for the simple adiabatic protocols with different ramp-down times ($t_{\rm ramp}$), whose field strength and interaction strength are shown in panel (b).
  (c) Double occupation $2n_d(t)$ for the sophisticated  protocols with different ramp-down times ($t_{\rm ramp}$), whose field strength and time-dependent interaction strength are shown in panel (d).
    Here, we choose $K_0=1.43$, and $\beta=0.30$.}
  \label{fig:AC_COMB_U175b03}
\end{figure}

In Fig.~\ref{fig:AC_COMB_U175b03}, we show the time evolution of the double occupation for different excitation protocols to check the adiabaticity. 
Figure~\ref{fig:AC_COMB_U175b03}(a) depicts the evolution under the simple protocol, see also Fig.~\ref{SI_fig_NR1_adiabaticity} below.
One can see that the final value of the double occupation is almost independent of the ramp-down time.
On the other hand, under the sophisticated protocol (panel (c)(d)), 
where the driving is ramped up away from resonance and the interaction is subsequently ramped into resonance (and vice versa for the ramp down), 
one can clearly see a reduction of the final value of the double occupation
as the ramp-down time of the interaction is increased.
These results are consistent with the experimental observation, where the sophisticated method shows a 
rather prominent reduction of the double occupation with increasing ramp-down time, see Fig.~\ref{SI_fig_NR_adiabaticity_2ramp}.

\subsection{Analysis of the single-particle spectrum}
\label{NDMFT_NR_spectrum}
\begin{figure}[t]
 \centering
   \hspace{-0.cm}
    \vspace{0.0cm}
	\includegraphics[width=60mm]{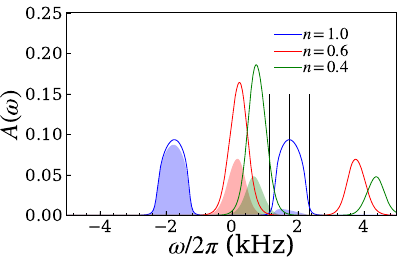}
  \caption{Local single-particle spectral function ($A(\omega)$) and its occupation ($N(\omega)$, shaded area) at $\beta=0.3$ in equilibrium for different fillings.
  The vertical lines indicate $\frac{U}{2}-\frac{W}{2},\frac{U}{2}$ and $\frac{U}{2}+\frac{W}{2}$.}
  \label{fig:fAkw_U175b1}
\end{figure}

In Fig.~\ref{fig:fAkw_U175b1}, we show the local single-particle spectral functions and corresponding occupations for different fillings at $U/h=3500$~Hz.
At half filling, one can observe clear Hubbard bands, whose centers are located near $\omega =\pm \frac{U}{2\hbar}$ and the bandwidth of each band is essentially $W$. We note that the band edges become sharper as temperature is decreased. 

Away from half-filling, with finite hole doping, the position of the lower band is located around $\omega=0$ and
the lower part of the lower Hubbard band is occupied.
Still the distance between the bands is almost $U$ and their bandwidth is similar.
The weight of the upper band is decreased from $\frac{1}{2}$, while that of the lower band is increased from $\frac{1}{2}$.
This is because the weight of the upper band, roughly speaking, reflects the probability of producing a doublon by adding a particle to the system. 
This probability is reduced if the system is hole doped.

Now we show that the effect of detuning mentioned above can be well understood by considering the structure of the spectral functions and occupations.
As can be seen in Eq.~(\ref{eq:chi_w}), the absorption ratio is associated with ${\rm tr}[\hat{\sigma}_2 \hat{A}_k(\omega+\Omega) \hat{\sigma}_2 \hat{A}_k(\omega)]$. 
In the present case, we numerically confirmed that the contribution from the off-diagonal component of $\hat{A}$ is much smaller than that from the diagonal terms.
Therefore, one can focus on $A_{{\bf k},AA}(\omega)=A_{{\bf k},BB}(\omega)$.
These spectral functions consist of two bands, which are separated by $U$ and have almost the same bandwidth as in $A(\omega)$. 
In particular, at half-filling, the shapes of the upper band and the lower band in $A_{{\bf k}}(\omega)$ (and hence in $A(\omega)$) turn out to be almost identical. 
Because of this, at half-filling, the lower band (occupied band) and the upper band (unoccupied band) overlap almost perfectly when they are shifted by $\hbar\Omega=U$
and this frequency shows the largest value of $|{\rm Im}\chi_{xx}(\Omega)|$. 
Positive and negative detuning can reduce the overlap.
However, because of the similar shapes of the upper and lower bands, the overlap is similar and the value of  ${\rm Im}\chi_{xx}(\Omega)$ also becomes similar for positive and negative detuning.
In the hole-doped case, the lower part of the lower band is occupied, and these states can be excited to the lower part of the upper band when $\hbar\Omega=U$.
For negative detuning $\hbar\Omega>U$, there exist final states in the upper band.
On the other hand, for positive detuning, the number of accessible final states is reduced because the occupation shifted by $\Omega$ may be located within the gap.

\begin{figure}[t]
 \centering
   \hspace{-0.cm}
    \vspace{0.0cm}
	\includegraphics[width=60mm]{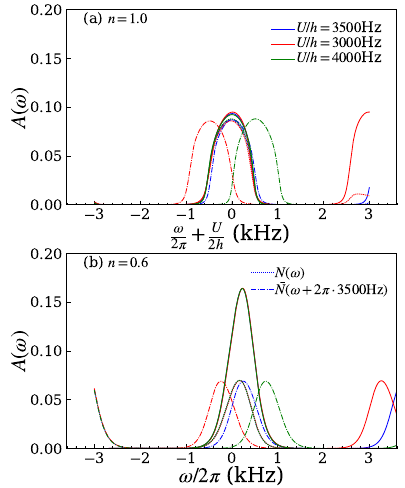}
  \caption{Local single-particle spectral function ($A(\omega)$), its occupation ($N(\omega)$, dashed lines)
   and the spectral functions shifted by indicated frequencies (dash-dotted lines) at $\beta=0.30$ in equilibrium. 
  Panel (a) is for $n=1.0$ and panel (b) is for $n=0.6$. The overlap between the occupation and the shifted spectrum for the unoccupied states roughly shows the efficiency of absorption under the excitation with the corresponding frequency.}
  \label{fig:Akw_U175b1_detune_excitation}
\end{figure}

We can directly demonstrate the above point using the local spectrum as a representative of $A_{\bf k}$.
In Fig.~\ref{fig:Akw_U175b1_detune_excitation} (a), we show the overlap between the occupation and the unoccupied states shifted by $\hbar\Omega$ for different $U$ at half filling.
One can see that the overlap is maximal at $U=\hbar\Omega$, and that for $U/h=3000$~Hz and $U/h=4000$~Hz it is  almost the same. 
Away from half-filling, one can identify clear differences in these overlaps between $U/h=3000$~Hz and $U/h=4000$~Hz. This originates from the larger occupation in the lower part of the lower band and the distance of the bands being almost $U$. 																		   
These observations naturally explain the effects of detuning on the speed of the doublon creation observed in Fig.~\ref{fig:AC_EXP_RAMP_U175b1_tr754_detune2_tmp}.

\section{Experiment}
\label{Exp}
\subsection{General preparation}
\label{Exp_preparation}
We start with a gas of $^{40}$K fermions in the $m_F=-9/2,-5/2$ sublevels of the $F=9/2$ manifold confined in a harmonic optical dipole trap. The mixture is evaporatively cooled down to a spin-balanced Fermi degenerate gas of 35(3)$\times 10^3$ atoms at a temperature $T/T_F=0.14(2)$ ($T_F$ is the Fermi temperature). Feshbach resonances (around 202.1 G and 224.2 G) allow us to tune the interactions from weakly to strongly repulsive using a $-9/2,-7/2$ or a $-9/2,-5/2$ mixture, respectively. The latter is prepared with a Landau-Zener transfer across the $-7/2\rightarrow-5/2$ spin resonance.

Four retro-reflected laser beams of wavelength $\lambda=1064$~nm are used to create the optical lattice potential
\begin{eqnarray} V(x,y,z) & = & -V_{\overline{X}}\cos^2(k
x+\theta/2)-V_{X} \cos^2(k
x)\nonumber\\
&&-V_{\widetilde{Y}} \cos^2(k y) -V_{Z} \cos^2(k z) \nonumber\\
&&-2\alpha \sqrt{V_{X}V_{Z}}\cos(k x)\cos(kz)\cos\varphi , 
\label{Lattice}
\end{eqnarray}
where $k=2\pi/\lambda$ and $x,y,z$ are the laboratory frame axes. The lattice depths $V_{\overline{X},X,\tilde{Y},Z}$ (in units of the recoil energy $E_R=h^2/2m\lambda^2$, $h$ is the Planck constant and $m$ the mass of the atoms) are independently calibrated using amplitude modulation on a $^{87}$Rb Bose-Einstein condensate (BEC). The visibility $\alpha=0.99(1)$ is calibrated using the same technique but in an interfering lattice configuration with $V_X,V_Z\not = 0$. The phases $\theta$ and $\varphi$ are set to $1.000(2)\pi$ and $0.00(3)\pi$, respectively and determine the lattice geometry. The experimental Hubbard parameters $J$ and $U$ are numerically calculated from the lattice potential's Wannier functions computed with band-projected position operators \cite{Uehlinger2013}. The bandwidth $W$ is defined as $W=2(J_x+2(J_y+J_z))$ for the single band tight-binding model. Tunneling terms $J_x,J_z$ are along the horizontal and vertical links of the hexagonal unit cell, and $J_y$ along the orthogonal direction of the hexagonal plane, see Fig.~\ref{figS3:HTS}. The tunneling link $J_w$ across the hexagonal unit cell is strongly suppressed. We note that the tunneling parameters $J_{\parallel},J_{\perp}$ introduced in the theoretical studies are related to the ones presented here for the experiments by  $J_{\parallel} = J_x$ and $J_{\perp}=J_y=J_z$. The estimated values for the Hubbard parameters used in the experiments, namely tunnelings $J_{x,y,z}$, bandwidth $W$ and on-site interaction $U$, are given in Tables \ref{Table_off-res} to \ref{Table_near-res-ad} for the different driving protocols used in the experiment, which were also used for the theoretical simulations.

\subsection{Periodic driving}
\label{Exp_driving}
The driving protocols are carried out with a piezoelectric actuator that periodically modulates in time the position of the $X$ retro-mirror at a frequency $\Omega/(2\pi)$ and amplitude $A$. As a result, the retro-reflected $X$ and $\overline{X}$ laser beams acquire a time-dependent phase shift with respect to the incoming beams and the lattice potential is $V(x-A\cos(\Omega t),y,z)$. The normalized amplitude is given by $K_0=mA\Omega d_x/\hbar$, with $d_x$ the inter-site distance along the $x$ axis ($\hbar=h/2\pi$). Values of $K_0$ are given in tables \ref{Table_off-res} to \ref{Table_near-res-ad} for all the driving protocols presented in this work. These include a factor of $d_x$ corresponding to the distance between Wannier functions located on the left and right sites of a lattice bond for the corresponding lattice geometry.

In addition, the phases of the incoming $X$ and $Z$ lattice beams are modulated at frequency $\Omega/(2\pi)$ using acousto-optical modulators to stabilize the phase $\varphi=0.00(3)\pi$. This compensation is not perfect and leads to a residual amplitude modulation of the lattice beams. Also, the phase and amplitude of the mirror displacement driven by the piezoelectric actuator is calibrated by measuring the amplitude of the phase modulation without any compensation.

\subsection{Detection methods}
\label{Exp_detection}
The detection of double occupancies starts with freezing the dynamics by ramping up the lattice depths to $V_{\overline{X},X,\tilde{Y},Z}=[30,0,40,30] \ E_R$ within $100 \ \mu$s. Freezing the lattice at different modulation times within one driving period allows us to average over the micromotion for each measurement protocol. Later, we linearly ramp off the periodic driving within 10 ms. Then, a (radio-frequency) Landau-Zener transition tuned to an interaction-dependent spin resonance selectively flips the $m_F=-7/2$ atoms on doubly occupied sites to the initially unpopulated $m_F=-5/2$ spin state (or $m_F=-5/2$ is flipped into the $m_F=-7/2$ state when we start with a $-9/2,-5/2$ mixture). Subsequently, we perform a Stern-Gerlach type measurement after switching off all confining potentials to separate the individual spin states. After 8 ms of ballistic expansion the $-9/2,-7/2,-5/2$ spin components are spatially resolved in an absorption image. For each spin component its spatial density distribution is fitted with a Gaussian profile to estimate the fraction of atoms on each spin state determining the fraction of double occupancies $\mathcal{D}$.

\subsection{Off-resonant modulation}
\label{Exp_OR}
Once the degenerate Fermi gas is prepared, the optical lattice is loaded by ramping up the lattice beams within 200 ms to an anisotropic configuration $J_{x,y,\tilde{z}}/h=[487,102,83]$~Hz (see table~\ref{table:Dens_profiles} and Fig.~\ref{figS3:HTS} for definition of $J_{\tilde{z}}=\tilde{J}_z=(2J_z+J_w)/3$). From this intermediate loading configuration we then ramp within 10 ms to a hexagonal configuration with reduced tunnelings $J_{x,y,z}/h=[191(31),41(3),40(3)]$~Hz (see table~\ref{Table_off-res}). The chosen values for the tunneling enables us to resolve dynamics on the order of the driving period. We leave the harmonic trapping potential and the anisotropy constant during the second loading process such that we get a locally equilibrated state.  
During this loading the interactions in a $m_F=-9/2,-7/2$ mixture are also set, see \emph{starting lattice} in Table \ref{Table_off-res}. In this configuration the system is prepared at $U/h=753(26)$ Hz and $W/h=707(65)$ Hz, right in the crossover regime from a metal to a Mott insulator. Then, the off-resonant driving protocol starts by ramping up $K_0$ from 0 to 1.70(2) within a variable amount of time $t_{\text{ramp}}$. The modulation is held very shortly (100 $\mu$s) before the detection of $\mathcal{D}$ starts. At ramp times below 100~$\mu$s (shaded area in Fig.~2) residual dynamics during the detection process and the finite bandwidth of the piezoelectric actuator influence the measured value of $\mathcal{D}$.
The level of $\mathcal{D}$ is compared to the one of an undriven (static counterpart) protocol. This static protocol consists in ramping down (up) the $V_{X}$ ($V_Z$) lattice depth (\emph{final lattice} configuration), which results in a ramp of $J_x/h$ from $193(34)$ Hz to $81(13)$ Hz. We note that this tunneling ramp does not exactly follow the 0th-order Bessel function tunneling renormalization $J^{\text{eff}}[t_{\text{ramp}}]\sim\mathcal{J}_0\left(K_0[t_{\text{ramp}}]\right)$ expected in the driven protocol. Further theoretical studies presented above (section \ref{NDMFT_NR_quench}) show that the differences in the dynamics are negligible between these two protocols. Two reference levels for $\mathcal{D}$ are additionally measured when we simply load the atoms in a static lattice of either $J_x/h=193(34)$~Hz (\emph{starting lattice}) or $J_x/h=81(13)$ Hz (\emph{final lattice}).

\begin{table}[t!]
\setlength\extrarowheight{5pt}
\begin{tabular}{|c || c | c|}
\multicolumn{3}{c}{\textbf{Off-resonant modulation}} \\
\firsthline
parameter & starting lattice & final lattice \\
\hline
\multicolumn{3}{c}{Driven system} \\
\hline
$J_{x,y,z}/h \left(\text{Hz}\right)$ & 191(31), 41(3), 40(3) & - \\
$W/h\left(\text{Hz}\right)$ & 707(65) & - \\
$J^{\text{eff}}_{x,y,z}/h \left(\text{Hz}\right)$ & - & 76(13), 41(3), 40(3) \\
$W^{\text{eff}}/h\left(\text{Hz}\right)$ & - & 478(32) \\
$U/h \left(\text{Hz}\right)$ & \multicolumn{2}{c|}{753(26)} \\
$K_0$ & \multicolumn{2}{c|}{1.70(2)} \\
\hline
\multicolumn{3}{c}{Undriven system} \\
\hline
$J_{x,y,z}/h \left(\text{Hz}\right)$ & 193(34), 41(3), 39(4) & 81(13), 41(3), 39(4) \\
$W/h\left(\text{Hz}\right)$ & 708(71) & 481(33) \\
$U/h \left(\text{Hz}\right)$ & \multicolumn{2}{c|}{752(27)} \\
$K_0$ & \multicolumn{2}{c|}{-} \\
\hline
\end{tabular}
\caption{Experimental parameters for the off-resonant modulation protocols in the driven and undriven cases. Errors in ($J_{x,y,z}$, $W$) and $K_0$ account for the uncertainty of the measured lattice depths $V_{\overline{X},X,\tilde{Y},Z}$ and uncertainty in the estimated lattice bond distance $d_x$, respectively. The error in $U$ additionally includes the uncertainty on the measured magnetic field and Feshbach resonance calibration. $J^{\text{eff}}_{x,y,z}$ (and so $W^{\textbf{eff}}$) are computed from the relation $J^{\textbf{eff}}=J\mathcal{J}_0(K_0)$.
}
	\label{Table_off-res}
\end{table}

\begin{figure}[bt]
    \includegraphics{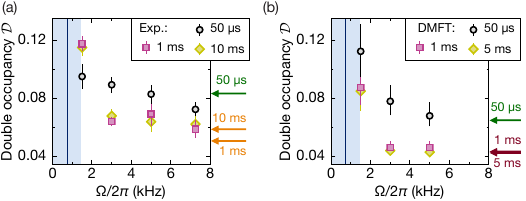}		
    \caption{Crossover from off-resonant to near-resonant modulation. Experimental (a) and DMFT (b) data display $\mathcal{D}$ measured after ramping up the drive for different frequencies. The vertical line indicates the interaction $U/h= 0.75(3)~\mathrm{kHz}$ and the blue area around it has the width of twice the bandwidth $W/h = 0.71(7)~\mathrm{kHz}$. The data is presented for three different ramp times corresponding to $50~\mathrm{\mu s}$ (circles), 1~ms (squares) and 10~ms (5~ms) (diamonds) for panel a (b). Arrows labelled with holding time indicate data taken in a static system. The upper green arrow shows the reference value holding in the initial ($J_x/h = 193(34) \mathrm{Hz}$) lattice for $50~\mathrm{\mu s}$. In the experimental data the two lower orange arrows denote the  reference value for the final lattice ($J_x/h = 81(13)~\mathrm{Hz}$) after holding 1~ms (lower value) and 10~ms (upper value). The lower red arrows in the DMFT data represent the value after a 1~ms and 5~ms ramp of the tunneling from initial to final. The $\mathcal{D}$ saturates at 1~ms and only a negligible change compared to 5~ms is seen. Error bars in (a) denote the standard error for 5 measurements and in (b) reflect the uncertainty of the entropy estimation in the experiment [39].}
	\label{SI_fig_OR_detuning}
\end{figure}

In Fig.~2 of the main text it is shown how the effective description breaks down when the drive frequency $\Omega/2\pi$ approaches the on-site interaction $U$. We present the same data in Fig.~\ref{SI_fig_OR_detuning} showing $\mathcal{D}$ as a function of drive frequency. The data displayed is taken after a ramp time of $50~\mathrm{\mu s}$ (circles), 1~ms (squares) and 10~ms for the experimental part (a) or 5~ms for the DMFT calculations (b) (diamonds). The solid vertical line denotes the value of the interaction and the blue area around it has the width of twice the bandwidth. In this region, which corresponds to the near-resonant driving regime, direct excitations of $\mathcal{D}$ are possible. The upper green arrow denotes the reference value taken after holding $50~\mathrm{\mu s}$ in the initial static lattice. The two lower orange arrows in panel (a) depict reference values in a static lattice with a tunneling of $J_x/h = 81(13)~\mathrm{Hz}$ after holding for 1~ms (lower) or 10~ms (upper). In panel (b) the lower red arrows correspond to the $\mathcal{D}$ reached after a ramp of the tunneling from the intial to the final value in 1~ms and 5~ms. The difference of these two values is negligible indicating the saturation in $\mathcal{D}$ after a 1~ms ramp.

The effective Hamiltonian predicts a change of $\mathcal{D}$ from the reference value in the initial lattice (arrow denoted with $50~\mathrm{\mu s}$) to the reference value of the final lattice (arrows denoted with longer hold times). For frequencies of 3~kHz and higher the measured and calculated $\mathcal{D}$ are within the expectation of the effective Hamiltonian whereas for 1.5~kHz a much larger value of $\mathcal{D}$ appears.
We attribute the remaining deviations in the values of $\mathcal{D}$ between the ab-initio calculations and the experimental values to the systematic uncertainties on the input temperature and density profiles provided by the experiment. 
Experimental data taken at $50~\mathrm{\mu s}$ is in addition influenced by residual dynamics during the detection and the finite bandwidth of the piezoelectric actuator. 

\subsection{Adiabaticity in off-resonant modulation}
\label{Exp_OR_adiab}
\begin{figure}[bt]
    \includegraphics{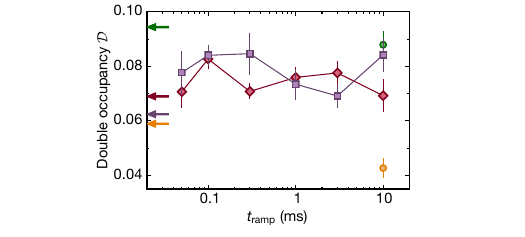}		
    \caption{Adiabaticity in off-resonant modulation. 
    {In the driven system with $\Omega/(2\pi)=7.25$ kHz (purple squares) $K_0$ is ramped up in 10 ms, held for 100~$\mu$s and then ramped down within a time $t_{\text{ramp}}$. In the undriven system (red diamonds) the lattice depths are directly ramped within the same time as for the driven case. 
Reference levels of $\mathcal{D}$ with $t_{\text{ramp}}=0$~s are shown (arrows).
For comparison, the level of $\mathcal{D}$ in the static systems with $J_x/h=79(12)$ Hz (yellow circle) and $J_x/h=208(34)$ (green circle) is shown when the waiting times are [10 ms + 100 $\mu$s] (arrows) or [10 ms + 100 $\mu$s + 10 ms] (circles). Data points are the mean and standard error of 5 individual runs (including an average over one driving period $(2\pi)/\Omega$ in the data versus time).}
	}\label{SI_fig_OR_adiabaticity}
\end{figure}

We perform tests on the adiabaticity to revert the suppression of $\mathcal{D}$ in the off-resonant protocols, see Fig.~\ref{SI_fig_OR_adiabaticity}. Here, the system is prepared in a driven lattice ($\Omega/(2\pi)=7.25$~kHz) where we use a 10 ms ramp up of $K_0$ to a final value of 1.68(2). From this \emph{starting lattice} configuration with $U/h=752(26)$~Hz and $J^{\text{eff}}_{x,y,z}/h = [85(15),42(4),38(4)]$~Hz (see Table~\ref{Table_off-res-ad}) we hold 100 $\mu$s before $K_0$ is ramped down to 0 within a variable amount of time $t_{\text{ramp}}$ reaching the \emph{final lattice} configuration where $J_x/h=208(37)$~Hz. The detected level of $\mathcal{D}$ is compared to an undriven protocol that ramps the lattice depths $V_X$ and $V_Z$ to emulate the tunneling renormalization in the driven protocol. This changes $J_{x,y,z}/h = [81(12),42(3),39(3)]$~Hz in the \emph{starting lattice} to $J_x/h=193(32)$~Hz in the \emph{final lattice} configuration. We observe that $\mathcal{D}$ reaches comparable values in the driven and undriven systems and we do not see a clear change versus ramp time. Reference levels of $\mathcal{D}$  were also measured when the system is prepared in a static lattice with $J_x/h=79(12)$ Hz or $J_x/h=208(34)$ Hz. For each of these measurements the level of $\mathcal{D}$ is detected [10 ms + 100 $\mu$s] or [10 ms + 100 $\mu$s + 10 ms] after the lattice loading is done. These waiting times simulate the time that it takes to ramp up (10 ms), hold (100 $\mu$s) and ramp down (10 ms) the driving amplitude $K_0$. We attribute the remaining dynamics to the adiabatically imperfect preparation.

\begin{table}[t!]
\setlength\extrarowheight{5pt}
\begin{tabular}{|c || c | c|}
\multicolumn{3}{c}{\textbf{Adiabaticity in off-resonant modulation}} \\
\firsthline
parameter & starting lattice & final lattice \\
\hline
\multicolumn{3}{c}{Driven system} \\
\hline
$J^{\text{eff}}_{x,y,z}/h \left(\text{Hz}\right)$ & 85(15), 42(4), 38(4) & - \\
$W/h\left(\text{Hz}\right)$ & 491(36) & - \\
$J_{x,y,z}/h \left(\text{Hz}\right)$ & - & 208(37), 42(4), 38(4) \\
$W^{\text{eff}}/h\left(\text{Hz}\right)$ & - & 737(76) \\
$U/h \left(\text{Hz}\right)$ & \multicolumn{2}{c|}{752(26)} \\
$K_0$ & \multicolumn{2}{c|}{1.68(2)} \\
\hline
\multicolumn{3}{c}{Undriven system} \\
\hline
$J_{x,y,z}/h \left(\text{Hz}\right)$ & 81(12), 42(3), 39(3) & 193(32), 42(3), 39(3) \\
$W/h\left(\text{Hz}\right)$ & 485(33) & 711(71) \\
$U/h \left(\text{Hz}\right)$ & \multicolumn{2}{c|}{750(27)} \\
$K_0$ & \multicolumn{2}{c|}{-} \\
\hline
\end{tabular}
\caption{Experimental parameters for the adiabaticity in off-resonant modulation protocols of the driven and undriven cases. Note that the parameters for the \emph{starting lattice} (\emph{final lattice}) configuration are the ones of the \emph{final lattice} (\emph{starting lattice}) configuration of Table \ref{Table_off-res}, because for the adiabaticity protocols we explore the ramping down of $K_0$. Each parameter value and error was computed as described in Table \ref{Table_off-res}.
}
	\label{Table_off-res-ad}
\end{table}

\subsection{Near-resonant modulation}
\label{Exp_NR}
In the near-resonant driving protocols used to study the creation of $\mathcal{D}$ versus $K_0$ and $U$ we load the cloud in 200 ms in an anisotropic lattice with $J_{x,y,\tilde{z}}/h=[202,100,67]$~Hz (see table~\ref{table:Dens_profiles}, and Fig.~\ref{figS3:HTS} for definition of $J_{\tilde{z}}=\tilde{J}_z=(2J_z+J_w)/3$). We further ramp in 10 ms to our starting hexagonal lattice configuration $J_{x,y,z}/h=[209(38),100(7),98(7)]$~Hz. In contrast to the off-resonant case we do not need to scale down the energy scales which results in a better preparation with respect to thermalization of the initial state. To reach strong repulsive interactions we use a $m_F=-9/2,-5/2$ mixture with its Feshbach resonance situated at 224.2 G. In particular we use $U/h=3.5(1)$~kHz to prepare the system in a deep Mott insulating state. The driving amplitude $K_0$ is then ramped up during a varying time and the modulation is held for 100 $\mu$s at constant $K_0$ before the detection of $\mathcal{D}$ starts. Here $\Omega/(2\pi)=3.5$~kHz in all protocols. Measurements with different $K_0$ are taken with fixed static Hubbard parameters, see Table \ref{Table_near-res}. Measurements with different $U$ are taken with fixed $K_0=1.44(2)$.

\begin{table}[t!]
\setlength\extrarowheight{5pt}
\begin{tabular}{|c || c|}
\multicolumn{2}{c}{\textbf{Near-resonant modulation}} \\
\firsthline
parameter & value(s) \\
\hline
\multicolumn{2}{c}{$\mathcal{D}$ versus $K_0$ protocol} \\
\hline
$J_{x,y,z}/h \left(\text{Hz}\right)$ & 209(38), 100(7), 98(7) \\
$W/h\left(\text{Hz}\right)$ & 1210(85) \\
$U/h \left(\text{kHz}\right)$ & 3.5(1) \\
$K_0$ & 0, 0.40(1), 0.90(1), 1.43(2), 1.90(2), 2.40(3) \\
\hline
\multicolumn{2}{c}{$\mathcal{D}$ versus $U$ protocol} \\
\hline
$J_{x,y,z}/h \left(\text{Hz}\right)$ & 195(35), 100(7), 102(8) \\
$W/h\left(\text{Hz}\right)$ & 1198(81) \\
$U/h \left(\text{Hz}\right)$ & 2.5(1), 3.0(1), 3.5(1), 4.0(1), 4.5(1) \\
$K_0$ & 1.44(2) \\
\hline
\end{tabular}
\caption{Experimental parameters for the near-resonant modulation protocols to measure $\mathcal{D}$ versus either $K_0$ or $U^{\text{eff}}$. The parameter values and errors were computed as described in Table \ref{Table_off-res}.
}
	\label{Table_near-res}
\end{table}

\begin{figure}[bt]
    \includegraphics{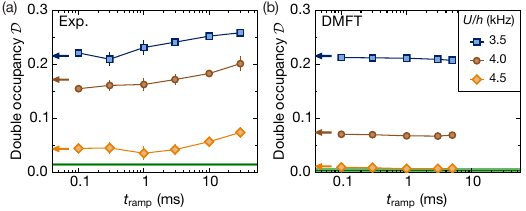}
    \caption{Adiabaticity in near-resonant modulation. 
    {Experimentally (a) and theoretically (b) obtained data measuring double occupancies $\mathcal{D}$ in a driven system ($\Omega/2\pi=3.5$ kHz) after ramping up $K_0$ from 0 to $1.44(2)$ within 10 ms, hold for 100 $\mu$s and then ramp $K_0$ down to $0$  in a strongly-interacting configuration with $J_{x,y,z}/h=[190(34),100(7),100(7)]$ Hz and $U/h=$ 3521(65) (blue squares), 4026(77) (brown circles) and 4531(90) Hz (orange diamonds). For each of these protocols, a reference level of $\mathcal{D}$ (blue, brown and orange arrows) is taken right after the 100 $\mu$s hold time. A reference level of $\mathcal{D}$ in the strongly-interacting static system is also shown (green line). No decay of $\mathcal{D}$ is observed. Error bars are shown for experimental
data indicating a standard error for 5 measurements.}
	}\label{SI_fig_NR1_adiabaticity}
\end{figure}			  
\subsection{Adiabaticity in near-resonant modulation}
\label{Exp_NR_adiab}
For the near-resonant driving protocols the adiabaticity to revert the creation of $\mathcal{D}$ is tested in two different measurement setups. In a first set of measurements the ramp up time of $K_0$ is fixed to 10 ms, the driving is held for another 100~$\mu$s at constant $K_0 = 1.44(2)$, and then $K_0$ is ramped down to 0. This is done for a varying $K_0$ ramp down time $t_{\text{ramp}}$, and the measurement is repeated at different interactions, see Table \ref{Table_near-res-ad}. Reference levels of $\mathcal{D}$ are taken right after the ramp up of $K_0$, i.e. before the ramp down, to compare the amount of $\mathcal{D}$ that is adiabatically reverted. We observe that $\mathcal{D}$ does not revert to the initial value for all ramp times $t_{\text{ramp}}$, i.e., with this ramp protocol, reaching the Floquet state on the timescales considered in this study is an irreversible process \cite{Goerg2018}. The DMFT simulations reproduce this non-adiabaticity for the same protocol, which makes it clear that the irreversibility is inherent to the driven system and not related to experimental imperfections.

\begin{figure}[bt]
    \includegraphics{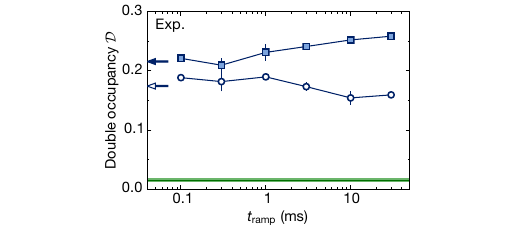} 		
    \caption{Adiabaticity in near-resonant modulation for different ramp protocols. 
    {In a lattice with $J_{x,y,z}/h=[200(50),99(9),99(11)]$~Hz and $U/h=3.5(1)$~kHz the modulation ($\Omega/(2\pi)=3.5$~kHz, $K_0=1.44(3)$) is ramped in 10~ms. We hold another 100~$\mu$s before $\mathcal{D}$ is measured for different ramp down times. We compare a two-step ramping scheme (empty circles) with a single-step ramping scheme (squares) which was shown already in Fig.~4. In the two-step protocol the modulation is ramped up and down first at $U_{\text{load}}/h=5.0(1)$~kHz away from resonance followed by a ramp of the interaction to $U_{\text{final}}/h=3.5$~kHz and back respectively.   
Reference levels of $\mathcal{D}$ before ramping back are shown (arrows) as well as in the undriven lattice (solid line).
Data points are the mean and standard error of 5 individual runs.}
	}\label{SI_fig_NR_adiabaticity_2ramp}
\end{figure}

In a second set of measurements a two-step ramp protocol is performed (Fig.~\ref{SI_fig_NR_adiabaticity_2ramp}), in the same way as it was done in \cite{Desbuquois2017}. Here we prepare the system at interaction $U_{\text{load}}/h=5.0(1)$~kHz and then switch-on the drive by ramping up $K_0$ from 0 to $1.43(2)$ within 10~ms. Afterwards, while driving, we ramp down the interaction to $U_{\text{final}}/h=3.5(1)$~kHz in another 10~ms. Then, the interaction is ramped back to $U_{\text{load}}$ within a variable amount of time $t_{\text{ramp}}$. Finally, $K_0$ is ramped down to 0 within another 10 ms before the detection of $\mathcal{D}$ starts. In addition, a reference level of $\mathcal{D}$ is taken right after the ramp down of the interaction from $U_{\text{final}}$ to $U_{\text{load}}$ to compare the amount of $\mathcal{D}$ that is adiabatically reverted with this protocol. In direct comparison to the single ramp protocol it can be seen that $\mathcal{D}$ decreases by around $10\%$. This relative decrease is very similar to the one seen in the theory data which was only taken for half filling (see Fig.~\ref{fig:AC_COMB_U175b03}). 

\begin{table}[t!]
\setlength\extrarowheight{5pt}
\begin{tabular}{|c || c|}
\multicolumn{2}{c}{\textbf{Adiabaticity in near-resonant modulation}} \\
\firsthline
parameter & value(s) \\
\hline
\multicolumn{2}{c}{ramp up $K_0$ protocol} \\
\hline
$J_{x,y,z}/h \left(\text{Hz}\right)$ & 190(34), 100(7), 100(7) \\
$W/h\left(\text{Hz}\right)$ & 1178(79) \\
$U/h \left(\text{Hz}\right)$ & 3521(65), 4026(77), 4531(90) \\
$K_0$ & 1.44(2) \\
\hline
\multicolumn{2}{c}{ramp up $U$ and $K_0$ protocol} \\
\hline
$J_{x,y,z}/h \left(\text{Hz}\right)$ & 211(38), 97(6), 97(8) \\
$W/h\left(\text{Hz}\right)$ & 1201(86) \\
$U_{\text{load}}/h \left(\text{Hz}\right)$ & 5032(106) \\
$U_{\text{final}}/h \left(\text{Hz}\right)$ & 3518(61) \\
$K_0$ & 1.43(2) \\
\hline
\end{tabular}
\caption{Experimental parameters for the near-resonant modulation protocols to measure $\mathcal{D}$ versus either $K_0$ or $U^{\text{eff}}$. Each parameter value and error were computed as described in Table \ref{Table_off-res}.
}
	\label{Table_near-res-ad}
\end{table}
			  
\subsection{Long-term dynamics in near-resonant modulation}
\label{Exp_NR_long}
We now comment on the experimental data for long-term dynamics $t_{\text{ramp}}\geq 10$~ms in the case of near-resonant driving, see Fig. 3. For the results with $U\approx\hbar\Omega$ and non-zero $K_0$ a decrease of double occupancy $\mathcal{D}$ is observed and no theoretical simulation comparison is provided in these cases. The underlying harmonic confinement can be described by a mean trapping frequency of $\overline{\omega}/(2\pi)=86$~Hz (see Table~\ref{table:Dens_profiles}) which gives us an estimation on the timescale $t_{\text{trap}}\simeq 10$~ms  above which trap induced dynamics can occur. Energy absorption from the drive can also lead to excitation of atoms to higher bands and shows up as unwanted heating in the lowest band models. An experimental indication of this heating is atom loss detected because atoms in higher bands can escape the trapped system. We only detect a significant atom loss on the order of 15\% at the last point for 300~ms ramp time. On the same very long timescale for this experiment we also expect technical heating arising mainly from fluctuations in the lattice depth. 

\begin{figure}[t]
	\includegraphics[width=1\columnwidth]{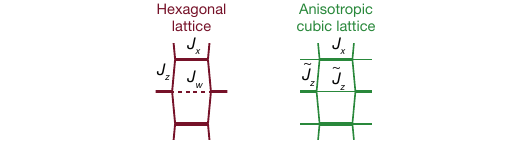}
	\caption{Illustrations of different lattice configurations. In the actual experiment, a hexagonal lattice configuration is used (left). To estimate the entropy and density profile in the harmonic trap, a high-temperature series calculation is performed for an anisotropic cubic lattice (right). To match the bandwidths of the two configurations, we use $\tilde{J}_z=(2J_z+J_w)/3$.
	}
	\label{figS3:HTS}
\end{figure}
\subsection{Density profiles}
\label{Exp_density}
In the experiment, the atoms are confined in a harmonic trapping potential. As a consequence, the density profile is inhomogeneous and all observables are averaged over the trap. In order to capture this in the numerical calculation, we determine the density profiles for our experimental parameters.

The main difficulty from an experimental point of view is to estimate the entropy of the atoms in the system. In order to determine it, we take additional reference measurements for two lattice configurations which are similar to the ones used in the off- and near-resonant driving regimes. For these systems, we measure three different observables:
\begin{itemize}
	\item The double occupancy $\mathcal{D}=2\left\langle \hat{n}_{i\uparrow}\hat{n}_{i\downarrow}\right\rangle$
	\item The transverse spin-spin correlator between neighboring sites in the $x$ direction $\mathcal{C}=-\bigl\langle \hat{S}^x_{i}\hat{S}^x_{i+1}\bigl\rangle-\bigl\langle \hat{S}^y_{i}\hat{S}^y_{i+1}\bigl\rangle$
	\item The nearest-neighbor correlator $\mathcal{N}=\left\langle \hat{n}_{i}\hat{n}_{i+1}\right\rangle$
\end{itemize}
($\left\langle ...\right\rangle$ denotes at the same time the expectation value and the trap average).

Afterwards, we run calculations based on a high-temperature series expansion \cite{Oitmaa2006} and try to reproduce the measured quantities by adjusting the entropy $S$ as a free parameter. All other parameters (Hubbard parameters, atom number and trapping frequencies) were derived from independent calibrations. The high-temperature series code is assuming an anisotropic cubic lattice geometry, which is very close to our hexagonal lattice configuration with the only difference that tunneling is allowed across the hexagonal unit cell (see Fig.\;\ref{figS3:HTS}). From this procedure, we obtain an estimate of the entropy per particle of $S/N=2.2(4)\;k_{\text{B}}$ and $S/N=2.4(2)\;k_{\text{B}}$ for the lattice configurations used in the off- and near-resonant driving regimes, respectively. 

\begin{table}[b]
\centering
\begin{tabular}{|c|c|c|c|}
\hline
\bf{Quantity} & \bf{Unit} & \bf{Off-resonant} & \bf{Near-resonant} \\
\hline
Atom number $N$ & $10^3$ & 35 & 35\\[2pt]
Entropy $S/N$ & $k_{\text{B}}$ & $2.2\pm 0.4$ & $2.4\pm 0.2$\\[2pt]
Tunnelings & \multirow{2}{*}{Hz} & \multirow{2}{*}{487,102,83} & \multirow{2}{*}{202,100,67}\\[2pt]
$(J_x,J_y,\tilde{J}_z)/h$ & & & \\[2pt]
Interaction $U/h$ & kHz & 1.1 & 3.5\\[2pt]
Trap frequencies & \multirow{2}{*}{Hz} & \multirow{2}{*}{66,65,140} & \multirow{2}{*}{64,80,123} \\[2pt]
$(\omega_x,\omega_y,\omega_z)/(2\pi)$ & & & \\[2pt]
Temperature $T$ & $J_x$ & $1.21^{+0.44}_{-0.36}$ & $3.31^{+0.57}_{-0.51}$\\[2pt]
Chem. potential $\mu$ & $J_x$& $1.42^{-0.74}_{+0.48}$ & $3.87^{-0.98}_{+0.80}$\\[2pt]
\hline
\end{tabular}
\caption{Parameters used to calculate the density profiles shown in Fig.\;\ref{figS1:Dens_profiles}. The entropy was estimated from reference measurements in similar lattice configurations. The tight-binding parameters and trap frequencies in the off-resonant regime are calculated for the intermediate lattice configuration, which sets the density profile. In the near-resonant regime, the parameters are the ones from the final lattice configuration (see Table\ref{Table_near-res}). In this case, the tight-binding parameters and mean trapping frequencies of the intermediate lattice configuration are matched to the final lattice. The upper and lower bounds on $T$ and $\mu$ correspond to the maximum and minimum value of the entropy given above.}
\label{table:Dens_profiles}
\end{table}

\begin{figure}[bt]
	\includegraphics[width=1\columnwidth]{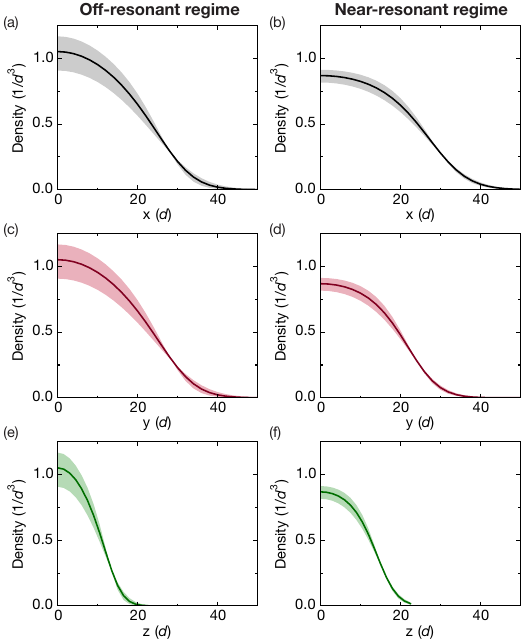}
	\caption{Density profiles used for the numerical calculations. Cut through the density profiles starting from the center of the harmonic trap in the $x$-, $y$ and $z$ directions for the off-resonant (a, c, e) and near-resonant modulation regimes (b, d, f). The lattice spacing is denoted as $d$. The profiles were calculated with a high-temperature series expansion for the parameters given in Table\;\ref{table:Dens_profiles}. The shading reflects the uncertainty of the density resulting from the error in temperature and chemical potential. This uncertainty is captured by the error bars shown for the numerical results in the figures of the main text. Both in the experiment and the numerical calculations the double occupancy is averaged over the inhomogeneous density profile of the cloud.
	}
	\label{figS1:Dens_profiles}
\end{figure}

We use this estimate of the entropies to calculate self-consistently the temperature $T$ and chemical potential $\mu$ with the same high-temperature series expansion. For this, we use the actual lattice configurations and trapping frequencies set in the experimental measurements. The results are summarized in Table\;\ref{table:Dens_profiles}. Finally, we use these parameters to calculate the density profile in the trap using a local density approximation $n(\mathbf{r})\equiv n(\mu-V_{\text{Trap}}(\mathbf{r}),T)= n(\mu-m/2 \sum{\omega_i^2 x_i^2},T)$, where $\mu$ is the chemical potential in the center of the trap and $\omega_i/(2\pi)$ are the trapping frequencies in the $x_i$-direction. The corresponding profiles are shown in Fig.\;\ref{figS1:Dens_profiles}. All numerical results shown in the main text are averages of the double occupancy over these inhomogeneous densities, with the error bars representing the uncertainties in the temperature and chemical potential.

\end{document}